\documentclass[11pt]{article}
\usepackage[letterpaper,margin=31mm]{geometry}
\usepackage[T1]{fontenc}
\usepackage{lmodern}
\usepackage{amsmath,amssymb,amsthm}
\usepackage{booktabs,longtable,array,calc,needspace}
\usepackage{microtype}
\usepackage{xurl}
\usepackage[hidelinks,bookmarksnumbered=false]{hyperref}
\newcommand{\tightlist}{\setlength{\itemsep}{0pt}\setlength{\parskip}{0pt}}
\newenvironment{paperresult}[2]
  {\par\Needspace{4\baselineskip}\addvspace{0.8\baselineskip}\noindent\hypertarget{#2}{}\textbf{#1.}\quad\itshape\ignorespaces}
  {\par\addvspace{0.65\baselineskip}}
\makeatletter
\renewcommand{\@maketitle}{%
  \begin{center}
    {\LARGE\bfseries\@title\par}%
    \ifx\@author\@empty\else\vskip .65em{\large\@author\par}\fi
  \end{center}
  \vskip .35em
}
\makeatother
\title{Sharp rank thresholds and cubic support in PPT extensions of completely symmetric states}
\date{}
\author{%
Qian Lilong\\[.15em]\small Independent Researcher\\[.15em]\normalsize\href{mailto:ssqianlilong@gmail.com}{\nolinkurl{ssqianlilong@gmail.com}}\\[.9em]
Teng Zhang\thanks{Corresponding author.}\\[.15em]\small Guangxi University, Nanning, Guangxi 530004, China\\[.15em]\normalsize\href{mailto:zhangteng@gxu.edu.cn}{\nolinkurl{zhangteng@gxu.edu.cn}}}
\hypersetup{pdfauthor={Qian Lilong, Teng Zhang},pdftitle={Sharp rank thresholds and cubic support in PPT extensions of completely symmetric states}}
\begin{document}
\maketitle
\begin{abstract}
For completely symmetric states on $d\ge3$ parties of actual local rank $N\ge3$, the least rank of an entangled state passing any prescribed finite positive level of the PPT extension hierarchy is $3d+N-3$. At equality, the two- and three-party marginals recover a ternary plane cubic shared by every extension. On this cubic the hierarchy becomes exact at a finite level precisely when every singularity outside the zero locus of the contraction quadratic is a split real node. We give the first exact level or an exact feasibility test for it, independent of $d$. On a smooth cubic the thresholds along the specified paths to the square of a flex tangent reduce to elliptic division and have unique rank-two Gram matrices. On reflection-invariant paths we obtain a quartic formula and the leading asymptotic constant. The least rank for entanglement itself is $3d+N-5$, and we classify equality. An explicit rational rank-nine three-qutrit state on a smooth cubic passes the first extension level and fails the second.
\end{abstract}
\smallskip
\noindent\textbf{2020 Mathematics Subject Classification.}
Primary 14P05. Secondary 13H10, 15A69, 81P40.

\noindent\textbf{Keywords.}
Completely symmetric states, PPT symmetric extensions, entanglement,
Hankel spectrahedra, real plane cubics.
\medskip

\section{1. Introduction}\label{s-1}

A completely symmetric (CS) state has entries invariant under every permutation of its \(2d\) bra and ket indices. Its actual local rank \(N\) is the rank of a one-party marginal. At level \(s\ge1\), one party is extended by \(s\) copies, the extension is invariant under permutations of those copies and the original party, and it is positive under partial transpose (PPT) across every bipartition. For a CS marginal, PPT across one copied single-party cut suffices and forces the extension itself to be CS.

\textbf{Sharp rank and reconstruction at equality.} For every prescribed finite \(s\ge1\), the least operator rank of an entangled CS state that passes level \(s\) is \(3d+N-3\) (\hyperlink{res-1A}{Theorem~1A}). The state attaining this rank may depend on \(s\), and passing all levels implies separability. Now fix an entangled state of that rank that passes level one. Its two-party marginal recovers the ternary local subspace and any complementary coherent directions, and its three-party marginal recovers the reduced plane cubic \(X\) supporting the ternary core of every extension (\hyperlink{res-1B}{Theorem~1B}). For three qutrits this means level one detects all rank-seven and rank-eight entangled CS states, while rank nine suffices to pass any prescribed finite level.

\textbf{Detection on the recovered support.} Partial trace acts on test forms on \(X\) by multiplication by a fixed positive definite quadratic \(g\). The first exact level of \((X,g)\) is the first level at which every entangled state on that support is detected. For a state as in \hyperlink{res-1B}{Theorem~1B}, this gives a uniform detection bound because its coherent blocks extend at every level. In \hyperref[s-1-3]{Section~1.3} the first detecting level of an explicit rational rank-nine state and the first exact level of its smooth support are both two.

\hyperlink{res-1D}{Theorem~1D} characterizes when the bound is finite. \hyperref[s-3-2]{Sections~3.2}--\hyperref[s-3-4]{3.4} give tests for its value, including elliptic division on smooth-cubic paths. Along a reflection-invariant path approaching the square of a flex tangent, the first exact level grows on the order of \(\epsilon^{-1/6}\) (\hyperlink{res-1H}{Corollary~1H}). \hyperref[s-1-1]{Sections~1.1} and \hyperref[s-1-2]{1.2} give two states on the same triangle, with identical marginal ranks and different detecting levels. The real filtering identity in \hyperref[contraction-and-real-local-filtering]{Section~2.1} identifies the hierarchy for \(g(x)=\|Ax\|^2\) on \(X\) with the standard partial-trace hierarchy on \(AX\).

\textbf{Relation to prior work.} Chen, Chu, Qian, and Shen \cite[Section VI]{cmp1} ask about separability of multi-qutrit CS states, the role of local ranks, and detection by symmetric extensions. Their real coherent-projector representation of separable CS states \cite[Theorems 14 and 16]{cmp1} gives the connection with positive moment functionals. Blekherman's rank bound is \(3d-2\), attained already in three variables \cite[Corollary 2.2]{cmp2}. At prescribed actual local rank \(N\), \hyperlink{res-1}{Theorem~1} gives the bound \(3d+N-5\) for entanglement and classifies the states attaining it. Theorems 1A and 1B establish the finite-level extension threshold and its equality structure for the hierarchy of Doherty, Parrilo, and Spedalieri \cite{cmp7}. Fawzi's nonrepresentability result concerns the unrestricted bipartite separable cone \cite{cmp17}. For curves, Scheiderer's stability and positive-section multiplier theorems give finiteness results \cite[Theorem 4.1 and Corollary 4.4]{cmp13} \cite[Theorem 4.11]{cmp18}. Plaumann treats unweighted sums of squares on reducible affine curves \cite[Theorem 3.7]{cmp14}, and \cite{cmp10} gives degree bounds when the multiplier may be chosen. Here partial trace fixes the multiplier to be \(g^s\), and the curve results determine or test the least admissible exponent uniformly over nonnegative forms.

Let \(S=\mathbb R[x_0,\ldots,x_{N-1}]\), with its standard grading, and take \(d\ge3\). Write \(L:S_{2d}\to\mathbb R\) for a nonzero functional satisfying \(L(p^2)\ge0\) for all \(p\in S_d\). Its middle matrix has entries \(L(x^{\alpha+\beta})\) in the monomial basis. The corresponding tensor operator has entries \(L(x_{i_1}\cdots x_{i_d}x_{j_1}\cdots x_{j_d})\) in the ordered tensor basis. The matrix in the normalized symmetric tensor basis is diagonally congruent to the middle matrix. Thus positivity and rank agree, and division by the positive trace gives a state. These are the completely symmetric (CS) states of Chen, Chu, Qian, and Shen \cite{cmp1}.

Put \(I=\operatorname{Ann}(L)\), where \(I_j=\{p:L(pS_{2d-j})=0\}\) for \(0\le j\le2d\) and \(I_j=S_j\) thereafter. The algebra \(A=S/I\) is Artinian Gorenstein of socle degree \(2d\) by the nondegenerate complementary-degree pairings. Write \(h_j=\dim A_j\). The \(j\)-party reduced matrix has bilinear form \(L(pqg^{d-j})\), where \(g=\sum_i x_i^2\). Its kernel is \(I_j\). Indeed, expansion of \(g^{d-j}\) as a positively weighted sum of monomial squares implies that a zero-norm \(p\) satisfies \(pS_{d-j}\subset I_d\). Multiplying by \(S_d\) then gives \(p\in I_j\). In particular \(h_1\) is the actual local dimension and \(h_d\) is the total rank. We may pass to the real actual local support and set \(N=h_1\).

\hyperref[s-2]{Section~2} proves the rank bounds and cubic reconstruction. The reconstruction uses the lower bound in \hyperlink{res-1A}{Theorem~1A} before the attainment assertion. \hyperref[s-3]{Section~3} proves the exactness criteria and then separates consecutive levels. \hyperref[s-A]{Appendix~A} gives an independent rational construction for attainment of the extension rank bound.

\subsection{1.1 A three-qutrit state that separates the first two levels}\label{s-1-1}

Let \(v_i=(1,1,1)+e_i\) for \(i=0,1,2\), and let \(y_i\) denote coordinates in this basis. These independent vectors have squared Euclidean norm six and pairwise inner product five. The projective triangle \(X\) on the \(v_i\) has equation \(q=y_0y_1y_2\). In these coordinates the Euclidean contraction quadratic is \(g=6\sum_i y_i^2+10\sum_{i<j}y_i y_j\). On the edge from \(v_i\) to \(v_{i+1}\), let \(p(x,y)=(x-y/2)(x-y)(x-2y)\), where \((x,y)\) are coefficients of the ordered endpoints. The restrictions \(p^2\) agree at every vertex, so they define a degree-six section \(f\) on \(X\), nonnegative on all its real points.

Use the following signed evaluation functional in degree eight, with indices modulo three. \[
M=
16\sum_{i=0}^{2}\sum_{r\in\{1/2,1,2\}}\operatorname{ev}_{r v_i+v_{i+1}}
+\frac{1}{81}\sum_{i=0}^2\operatorname{ev}_{v_i-v_{i+1}}
-\frac{2}{3}\sum_{i=0}^2\operatorname{ev}_{v_i}.
\] Evaluating a homogeneous section at a physical-space vector means substituting its \(y\) coordinates. Let \(T=M(g^4)=8435166433/27\) and \(L(f')=M(gf')/T\). If \(M=\sum_k w_k\operatorname{ev}_{z_k}\) as above, the density matrix in the standard tensor basis is \[
\rho=\frac1T\sum_k w_k\|z_k\|^2
   z_k^{\otimes3}(z_k^{\otimes3})^{\mathsf T}.
\] The four-party extension uses coefficients \(w_k\) and vectors \(z_k^{\otimes4}\) with the same normalizing constant \(T\).

In quotient monomial bases omitting multiples of \(q\), the moment matrices with entries \(M(g^{4-j}mp)\) for degree-\(j\) monomials \(m,p\) are positive definite. The accompanying verifier checks all Sylvester pivots over \(\mathbb Q\).

\begin{longtable}[]{@{}rrl@{}}
\toprule\noalign{}
Parties \(j\) & Rank & Ambient homogeneous kernel \\
\midrule\noalign{}
\endhead
\bottomrule\noalign{}
\endlastfoot
1 & 3 & \(0\) \\
2 & 6 & \(0\) \\
3 & 9 & \(\mathbb R q\) \\
4 & 12 & \(q\,\mathbb R[y_0,y_1,y_2]_1\) \\
\end{longtable}

The three-party kernel recovers the triangle, and the positive four-party CS extension certifies passage at level one. Since \(L(f)=-6/T<0\), the state is entangled. Any separable representation by real product rays would be supported on \(X\), by \(L(q^2)=0\), and would have \(L(f)\ge0\).

For an obstruction at level two let \(c=5/6\). Define degree-five sections \(\Phi_i\) by their restrictions to each oriented cyclic edge \(ij\). The two nonzero restrictions there are \(\Phi_i=p(x,y)(x^2+xy/c)\) and \(\Phi_j=-p(x,y)(y^2+xy/c)\). The restrictions agree at the vertices. Let \(K\) have diagonal entries one and off-diagonal entries \(-7/18\), so its eigenvalues are \(2/9,25/18,25/18\). Expansion on the edges gives \[
g^2 f=36\,\Phi^{\mathsf T}K\Phi
\qquad\text{on }X .
\] Any extension at level two would share the degree-three kernel \(q\), hence be supported on \(X\) by the marginal kernel identity. Its positivity on this sum of squares would force \(L(f)\ge0\), proving failure at level two.

The separation also holds on a full-rank open set of completely symmetric states. \hyperref[s-2-3]{Section~2.3} gives an ambient witness and an explicit interval of isotropic separable noise for this example.

\subsection{1.2 The same support at two detection levels}\label{s-1-2}

We give a rational state on the same triangle with marginal ranks \((3,6,9)\). It already fails level one. Work in the vertex coordinates of \hyperref[s-1-1]{Section~1.1} and put \(R=\mathbb R[y_0,y_1,y_2]/(q)\). On every oriented edge write its coordinates as \((t,u)\), put \(c=5/6\), and write \(g_0=g/6\), whose restriction is \(t^2+2ctu+u^2\). Define \[
p_0(t,u)=(t-u/2)(t-2u),\qquad P(t,u)=(t+u)p_0(t,u).
\] Both polynomials take the value one at each endpoint. Let \(Z\) consist of the nine points with \(u=1\) and \(t\in\{-1,1/2,2\}\) on the three cyclic edges, and distinguish the point \(b\) at \(t=-1\) on the edge \(20\). In degree six set \[
L_0=\sum_{z\in Z\setminus\{b\}}\operatorname{ev}_z
       -\frac4{203}\operatorname{ev}_b,\qquad
L_-=L_0+\frac1{100}\bigl(\operatorname{ev}_{v_0}+\operatorname{ev}_{v_1}\bigr).
\]

\textbf{Positivity and rank.} A degree-three section on \(X\) vanishing at all of \(Z\) must restrict on each edge to a multiple of \(P\). Matching the endpoint values makes these multiples equal. Thus the evaluation map on \(R_3\) has rank eight. For a cubic section \(a\), subtract its leading coefficient times \(P(t,1)\) on each edge and interpolate the resulting quadratic at the three roots of \(P\). Its value at zero is the end value of \(a\) minus the start value, and the interpolation weights there are \((2,8,-1)/9\). Summing the three endpoint differences gives \(\sum_{z\in Z} k_z a(z)=0\), where \(k_z\) takes the values \(2,8,-1\) in the given root order. In particular \(k_b=2\), and the eight evaluations away from \(b\) are independent. The squared norm of their coefficient vector \((k_z)_{z\ne b}\) is \(203\). Consequently \[
L_0(a^2)=\sum_{z\ne b}a(z)^2-\frac{1}{203}
                \left(\sum_{z\ne b}k_z a(z)\right)^2 .
\] This is positive semidefinite of rank seven. In its kernel, the eight values have the form \(a(z)=\lambda k_z\). On the edge \(01\) the end value minus the start value is \(69\lambda/9\). Therefore if \(a(v_0)=a(v_1)=0\), then \(\lambda=0\) and \(a\) is a multiple of the section with restrictions \(P\). That multiple is zero as well. Adding the two vertex evaluations makes \(L_-\) positive definite on \(R_3\). Normalization by \(L_-(g^3)>0\) gives a CS state of rank nine, with cubic kernel \(\mathbb R q\) and marginal ranks \((3,6,9)\) by the marginal-kernel identity.

\Needspace{6\baselineskip}
\textbf{Failure at level one.} Let \(f_-\) restrict to \(P^2\) on edges \(01\) and \(12\), and to \(p_0^2g_0\) on edge \(20\). The restrictions agree at the vertices. The form is nonnegative on \(X\), vanishes at the eight positive nodes, and has value \(27/4\) at \(b\). Thus \(L_-(f_-)=-1147/10150\). To certify failure at level one, it suffices to show that \(g_0 f_-\) is a sum of squares on \(X\). On the first two edges use \(P\) times the linear vector \(t v_i/\sqrt6+u v_{i+1}/\sqrt6\). On edge \(20\) use \(p_0\) times a quadratic vector with squared norm \(g_0^2\). In the basis \((t^2,tu,u^2)\) one Gram matrix for this quadratic vector is \[
\begin{pmatrix}
1&5/3&5/6\\
5/3&28/9&5/3\\
5/6&5/3&1
\end{pmatrix}.
\] Its leading principal minors are \(1\), \(1/3\) and \(2/81\). The vector at the two endpoints has unit norms and inner product \(c\), so an orthogonal change sends these two vectors to \(v_2/\sqrt6\) and \(v_0/\sqrt6\). The three edge vectors now agree at the vertices and their entries descend to degree-four sections on \(X\). This gives the claimed sum of squares. A CS extension would inherit \(q\) from the marginal and would therefore make \(L_-(f_-)\ge0\), a contradiction. Since every fully separable CS state passes level one, the new state is entangled. The state in \hyperref[s-1-1]{Section~1.1} has the same ranks, support and contraction metric and passes level one.

\subsection{1.3 An explicit rank-nine state on a smooth cubic}\label{s-1-3}

Let \(E\) be the cubic \(y^2z=x^3+xz^2\), and put \(g=x^2+y^2+64z^2\) and \(f=xz(x-z)^4\). On the real affine chart \(z=1\) one has \(x\ge0\), so \(f\ge0\) on \(E(\mathbb R)\), including at infinity. Write \(D(x)=x^3+x\) and \(G(x)=D(x)+x^2+64\). For a form \(P\) of degree eight on \(E\), reduce \(P(x,y,1)\) to \(U(x)+yV(x)\) using \(y^2=D(x)\). Define \(M(P)=\sum_r w_r U(r)\) with the following rational nodes and weights.

\begin{longtable}[]{@{}
  >{\raggedright\arraybackslash}p{(\columnwidth - 14\tabcolsep) * \real{0.0968}}
  >{\raggedleft\arraybackslash}p{(\columnwidth - 14\tabcolsep) * \real{0.1290}}
  >{\raggedleft\arraybackslash}p{(\columnwidth - 14\tabcolsep) * \real{0.1290}}
  >{\raggedleft\arraybackslash}p{(\columnwidth - 14\tabcolsep) * \real{0.1290}}
  >{\raggedleft\arraybackslash}p{(\columnwidth - 14\tabcolsep) * \real{0.1290}}
  >{\raggedleft\arraybackslash}p{(\columnwidth - 14\tabcolsep) * \real{0.1290}}
  >{\raggedleft\arraybackslash}p{(\columnwidth - 14\tabcolsep) * \real{0.1290}}
  >{\raggedleft\arraybackslash}p{(\columnwidth - 14\tabcolsep) * \real{0.1290}}@{}}
\toprule\noalign{}
\begin{minipage}[b]{\linewidth}\raggedright
\(r\)
\end{minipage} & \begin{minipage}[b]{\linewidth}\raggedleft
\(-1/10\)
\end{minipage} & \begin{minipage}[b]{\linewidth}\raggedleft
\(0\)
\end{minipage} & \begin{minipage}[b]{\linewidth}\raggedleft
\(9/10\)
\end{minipage} & \begin{minipage}[b]{\linewidth}\raggedleft
\(1\)
\end{minipage} & \begin{minipage}[b]{\linewidth}\raggedleft
\(2\)
\end{minipage} & \begin{minipage}[b]{\linewidth}\raggedleft
\(15\)
\end{minipage} & \begin{minipage}[b]{\linewidth}\raggedleft
\(30\)
\end{minipage} \\
\midrule\noalign{}
\endhead
\bottomrule\noalign{}
\endlastfoot
\(w_r\) & \(1\) & \(1\) & \(250\) & \(400\) & \(1/32\) & \(1/(15\cdot10^8)\) & \(1/(125\cdot10^{10})\) \\
\end{longtable}

This formula defines a real functional by reducing in the coordinate ring. The negative node enters only through the even part \(U\).

\textbf{Positive four-party extension.} A quartic on \(E\) has affine normal form \(A(x)+yB(x)\) with \(\deg A\le6\) and \(\deg B\le4\), by the pole orders of \(x\) and \(y\) at the flex. Its square has value under \(M\) equal to \(\sum_r w_r(A(r)^2+D(r)B(r)^2)\). The seven distinct nodes make the first part positive definite in \(A\). For the second part let \(r_0=-1/10\) and let \(J=\{9/10,1,2,15,30\}\). For \(r\in J\) put \[
\lambda_r=\prod_{\substack{t\in J\\t\ne r}}\frac{r_0-t}{r-t}.
\] Interpolation gives \(B(r_0)=\sum_{r\in J}\lambda_r B(r)\). Weighted Cauchy--Schwarz bounds the negative contribution at \(r_0\) by \(\theta\) times the sum over \(J\), where \[
\theta=-D(r_0)\sum_{r\in J}\frac{\lambda_r^2}{w_rD(r)}<\frac9{10}.
\] The displayed inequality is rational. It makes the second part positive definite on polynomials of degree at most four. Thus \(M\) is positive definite on \(R_4\), where \(R\) is the homogeneous coordinate ring of \(E\).

Set \(T=M(g^4)>0\) and \(L(P)=M(gP)/T\) for sextics. In standard Euclidean coordinates take \(v=(x,y,8z)\) and, for \(\mathbf i,\mathbf j\in\{1,2,3\}^3\), set \[
\rho_{\mathbf i,\mathbf j}
=\frac{M\bigl(g\,\prod_{k=1}^3 v_{i_k}v_{j_k}\bigr)}{T}.
\] The analogous four-party matrix with four factors on each side and no multiplier \(g\) is a positive extension with trace one. Contraction gives \(\rho\) since \(\sum_i v_i^2=g\). All entries are rational. Positive definiteness on \(R_4\), together with the marginal-kernel identity, gives the ranks \((3,6,9,12)\) of the first four marginals. In the physical coordinates \((v_1,v_2,v_3)\), the cubic kernel is generated by \(8v_2^2v_3-64v_1^3-v_1v_3^2\).

\textbf{The first exact level on the smooth support.} In the affine ring, set \(b=(1,x,x^2)^{\mathsf T}\) and take
\[
U=\begin{pmatrix}2559&-2228&230\\-2228&2138&-487/2\\230&-487/2&39\end{pmatrix},
\qquad
V=\begin{pmatrix}4096&-2431/2&63\\-2431/2&363&-37/2\\63&-37/2&1\end{pmatrix}.
\]
Direct coefficient comparison gives \(G^2=x\,b^{\mathsf T}Ub+(1+x^2)b^{\mathsf T}Vb\). The leading principal minors are \((2559,507158,18033497/4)\) for \(U\) and \((4096,37631/4,541/4)\) for \(V\). Using \(y^2=x(1+x^2)\), multiplication by \(x\) and homogenization give a positive definite Gram matrix for \(g^2xz\) on the six-dimensional reference space of Theorem~1F, with bases \((xz^2,x^2z,x^3)\) and \((yz^2,xyz,x^2y)\). Thus level two is exact for the entire support, for every \(d\ge3\). The level-one Gram matrix \(Q_1(g)\) in Theorem~1F has a principal minor \(1-(63/2)^2=-3965/4\), so the first exact level on \((E,g)\) is two. Both strict signs persist under small changes of the contraction quadratic on \(E\). In level two this uses surjectivity of the Gram map after division by \(xz\), by normal generation of the degree-six line bundle in \hyperref[s-3-3]{Section~3.3}.

\textbf{Detection of the explicit state.} Multiplying the SOS for \(g^2xz\) by \((x-z)^4\) gives an SOS for \(g^2 f\) on \(E\). By \hyperlink{support-prop}{Proposition~3.1} any level-two extension of \(\rho\) must obey the same cubic equation, so its contraction would satisfy \(L(f)\ge0\). Direct substitution gives
\[
M(gf)=-\frac{8582215363}{7812500000}<0.
\]
This excludes the extension and proves entanglement. The supplement checks the interpolation bound, the ranks, and the two level tests with rational arithmetic.

\section{2. Positivity and linear sections}\label{s-2}

The bilinear forms \(L(pqg^{d-j})\) induce positive definite forms on \(A_j\) for \(j\le d\). We also use the following consequence. If \(u^2\in I\) for a real homogeneous \(u\) of degree at most \(d\), then \(u\in I\). Apply positivity to each \((up)^2\) with \(p\in S_{d-\deg u}\) and use the middle kernel. If a real linear form \(u\) satisfies \(u^m\in I\) with \(m\le2d\), increase an odd exponent to the next even exponent and apply this argument repeatedly. This yields \(u\in I_1\).

\textbf{Generic multiplication in a subspace.} Let \(U\subset A_s\) be a real linear subspace, with \(s\ge1\), \(j\ge0\), and \(j+s\le d\). For a general \(u\in U\), \[
\ker(u:A_j\longrightarrow A_{j+s})=\{p\in A_j:Up=0\}.
\] Indeed, at a point of maximal rank, any kernel vector \(p\) extends along \(u+tv\) to a differentiable family \(p(t)\) of kernel vectors, for each \(v\in U\). Differentiation gives \(vp+up'(0)=0\) in \(A_{j+s}\). Multiply by \(vp\,g^{d-j-s}\) and apply \(L\). The term containing \(up'(0)\) vanishes by commutativity and \(up=0\), so \(L((vp)^2g^{d-j-s})=0\). The positive definite pairing on \(A_{j+s}\) gives \(vp=0\).

For \(U=A_s\) this common kernel is zero. Indeed \(A_{2d-j}=A_s A_{2d-j-s}\), and the perfect pairing against \(A_{2d-j}\) detects every nonzero \(p\in A_j\). Thus a general degree-\(s\) multiplier is injective wherever its target degree is at most \(d\). In particular a general linear form \(\ell\) is strong Lefschetz: \(\ell^{d-j}\) is injective into \(A_d\), so \(L(\ell^{2d-2j}pq)\) is positive definite on \(A_j\) and \(\ell^{2d-2j}:A_j\to A_{2d-j}\) is an isomorphism by duality.

\textbf{Extremality and base points.} The minimal face through \(L\) consists of positive middle matrices whose kernel contains \(I_d\). Its linear span is the annihilator of \(I_d S_d\), since \(L\) is positive definite on its middle-matrix range and permits sufficiently small perturbations in all those directions. Consequently extremality is equivalent to codimension one for \(I_d S_d\subset S_{2d}\). If a complex projective point \(z\) were a base point of \(I_d\), evaluation at \(z\) would belong to this annihilator after complexification. It would then be proportional to \(L\). Evaluation has middle rank one, ruling this out for a non-point ray. This is the standard Hankel face criterion \cite[3]{cmp2}.

\textbf{The two-step rank gap.} On the \(1\mid(d-1)\) cut, the CS cone is a face of the PPT cone. If \(P\) is the projection onto the symmetric tensor subspace and \(Q=1-P\), the equality face of the nonnegative functional \(X\mapsto\operatorname{tr}(QX)+\operatorname{tr}(QX^{\Gamma})\) imposes symmetric support on both \(X\) and \(X^\Gamma\). Symmetric support gives invariance under permutations of the row slots and, independently, of the column slots. After undoing the single-party transpose, the same invariance for \(X^\Gamma\) gives a transposition between the first column slot and a row slot from another party. Together these transpositions generate the full permutation group on the \(2d\) slots. This gives complete symmetry. The argument applies on the real actual support on each side of the cut. When \(h_1\ge4\), generic multiplication gives \(h_{d-1}\ge h_1\). The low-rank PPT theorem \cite{cmp4} excludes entangled states of rank at most \(h_{d-1}\) on this cut. Chen--Đoković \cite[Theorem 43]{cmp5} excludes extremality at rank \(h_{d-1}+1\) since both cut ranks exceed three. A non-point CS extreme ray is entangled on the cut by \cite[Theorem 16]{cmp1} and is PPT extreme because the CS cone is a face. Thus \(h_d-h_{d-1}\ge2\).

\hypertarget{binary-minimal-growth}{}\textbf{Minimal growth in two variables.} For \(\mathbb K=\mathbb R\) or \(\mathbb C\), let \(V\subset\mathbb K[x,y]_m\) have dimension \(r>0\). Then \(\dim(xV+yV)\ge r+1\), with equality only if \(V=q\mathbb K[x,y]_{r-1}\) for a form \(q\) of degree \(m-r+1\). The inequality holds because the largest power of \(x\) dividing every element of \(xV\) is one higher than for \(yV\), so the two \(r\)-dimensional spaces are distinct. For equality, proceed by induction on \(r\). The case \(r=1\) is immediate. When \(r>1\), set \(W=\{p:xp,yp\in V\}\) in degree \(m-1\). The intersection \(xV\cap yV=xyW\) gives \(\dim W=r-1\). Since \(xW+yW\subset V\), the inequality is an equality for \(W\), and induction gives \(W=q\mathbb K[x,y]_{r-2}\). Thus \(V=xW+yW=q\mathbb K[x,y]_{r-1}\).

\Needspace{5\baselineskip}
\hypertarget{small-hilbert-bound}{}\textbf{Small Hilbert values and base points.} Let \(K\) be a homogeneous ideal in a complex polynomial ring \(T\). If \(m\ge2\) and \(r=\dim(T/K)_m\le2\), then \(\dim(T/K)_{m+1}\le r\). If equality holds with \(r>0\), the forms in \(K_{m+1}\) have a common projective zero.

Identify the duals of \((T/K)_m\) and \((T/K)_{m+1}\) with spaces \(P,W\) of symmetric tensors. Multiplication by a linear form is dual to first contraction, so every first contraction of an element of \(W\) belongs to \(P\). If \(r=0\), then \(W=0\). A nonzero form whose first contractions span a space of dimension one is a scalar multiple of a pure power: choose coordinates in which all but one of its coordinate derivatives vanish. If every nonzero element of \(W\) has this property, then \(\dim W\le1\), since a sum of two independent pure powers has two independent first derivatives. A nonzero pure power in \(W\) is proportional to a point evaluation and supplies the claimed base point when \(\dim W=r\).

In the remaining case some \(F\in W\) has two independent first contractions, which span \(P\). The kernel of the contraction map has codimension two, so \(F\) uses just two variables after a linear change \cite[Proposition 1]{cmp6}. Write \(H\) for their two-dimensional space of linear tensors. Then \(P\subset\operatorname{Sym}^m H\). Since all first contractions of any tensor in \(W\) lie in \(\operatorname{Sym}^m H\), each tensor has its last \(m\) slots in \(H\). By symmetry this holds in every slot, so the tensor belongs to \(\operatorname{Sym}^{m+1}H\). Let \(U\) be the annihilator of \(P\) in the binary forms \(\operatorname{Sym}^m H^*\), so \(\dim U=m-1\). The space \(W\) annihilates \(H^*U\). Binary minimal growth gives \(\dim(H^*U)\ge m\), hence \(\dim W\le2\). If \(\dim W=2\), equality in binary growth gives \(U=q\operatorname{Sym}^{m-2}H^*\) for a quadratic \(q\). Now \(W\) is the entire annihilator of \(H^*U\). Evaluation at a complex projective zero of \(q\) belongs to \(W\) and gives the desired base point.

\textbf{Application to the middle degree.} Suppose \(L\) spans a non-point extreme ray, so \(I_d\) has no complex projective base point. Let \(\delta_j=h_j-h_{j-1}\). For a general linear form \(\ell\), \(A/(\ell)\) has Hilbert values \(\delta_j\) through degree \(d\). Apply the preceding argument to the complexified section. Starting at degree two, a Hilbert value of at most two cannot increase. If \(d\ge3\), \(0<\delta_{d-1}\le2\), and \(\delta_d=\delta_{d-1}\), the middle section has a projective base point, contradicting the absence of complex base points of \(I_d\). Thus the inequality at the middle degree is strict when \(0<\delta_{d-1}\le2\). For \(h_1\ge4\), the two-step rank gap gives \(\delta_d\ge2\), so \(\delta_j\ge3\) for every \(2\le j<d\).

\hypertarget{section-lifting}{}\textbf{Lemma 2.1 (Lifting regularity from a linear section).} Let \(J\subset S\) be a homogeneous ideal generated in degrees at most \(k\), with \(k\ge1\). Suppose multiplication by a linear form \(\ell\ne0\) is injective from degree \(j\) to degree \(j+1\) in \(S/J\) for \(0\le j\le k\), and the image ideal \(\bar J\) in \(S/(\ell)\) is \(k\)-regular. Then \(\ell\) is a nonzero divisor on \(S/J\), and \(J\) is saturated and \(k\)-regular.

Set \(T=S/(\ell)\). The natural surjection \(J/\ell J\to\bar J\) has degree-\(m\) kernel \((J_m\cap\ell S_{m-1})/\ell J_{m-1}\). By the assumed injectivity, this kernel is zero for \(m\le k+1\). Choose homogeneous lifts of the minimal generators of \(\bar J\) through this surjection. They generate \(J/\ell J\), since they span through degree \(k\) and \(J/\ell J\) is generated in those degrees. Let \(E\) be the graded free \(T\)-module on these generators. The kernel of \(E\to\bar J\) is generated in degrees at most \(k+1\) by \(k\)-regularity. Every such generator maps to zero in \(J/\ell J\), by the same low-degree isomorphism. Thus \(E\to J/\ell J\) factors through \(\bar J\), giving an inverse to the surjection \(J/\ell J\to\bar J\).

Its kernel vanishes in all degrees, so \((J:\ell)=J\). Also \(J\otimes_S T=\bar J\) and \(\operatorname{Tor}_i^S(J,T)=0\) for \(i>0\), using the one-step resolution of \(T\) and injectivity of \(\ell\) on \(J\subset S\). A minimal free resolution of \(J\) thus stays exact and minimal modulo \(\ell\), giving \(k\)-regularity of \(J\). Finally \(S/J\) has no irrelevant-ideal torsion because \(\ell\) is a nonzero divisor, so \(J\) is saturated.

\Needspace{5\baselineskip}
\hypertarget{associated-points}{}\textbf{Lemma 2.2 (Excluding zero-dimensional associated points).} Let \(L\) be a nonzero functional of degree \(2d\) with apolar ideal \(I\), and assume \(I_d\) has no complex projective base point. Suppose \(J\subset I\) is saturated, \(J_k=I_k\) for some \(1\le k\le d\), and the maps from forms to sections on \(X=\operatorname{Proj}(S/J)\) are surjective in degrees \(k,d\). Then \(X\) has no isolated or embedded zero-dimensional associated points. The surjectivity follows in particular if \(J\) is \(k\)-regular, by the vanishing of \(H^1(\mathbb P^{N-1},\widetilde J(t))\) for \(t\ge k-1\).

\textbf{Proof.} If \(X\) has a zero-dimensional associated point, the largest ideal sheaf \(\mathcal T\subset\mathcal O_X\) with finite support is nonzero. Its twist \(\mathcal T(k)\) has a nonzero global section because it is a coherent sheaf on a finite scheme. Lift the section to a form \(p\in S_k\), so \(p\notin J_k\). Choose \(F\in I_d\) nonvanishing at every geometric point of the support of \(\mathcal T\), using the absence of base points and avoiding a finite union of proper real linear subspaces. Multiplication by \(F\) gives an isomorphism \(\mathcal T(d)\to\mathcal T(2d)\).

For any \(u\in S_{2d-k}\), the section of \(pu\) lies in \(H^0(X,\mathcal T(2d))\). Divide it by \(F\) and lift the resulting degree-\(d\) section to a form \(b\in S_d\). Then \(pu-Fb\in J_{2d}\) by saturation. Since \(J\subset I\) and \(F\in I_d\), we have \(L(pu)=0\). Varying \(u\) gives \(p\in I_k=J_k\), a contradiction.

\Needspace{5\baselineskip}
\hypertarget{ternary-low-degree}{}\textbf{Low-degree equations in three variables.} Suppose \(S=\mathbb R[x_0,x_1,x_2]\) and \(I_D\) has no complex projective base point for a nonzero degree-\(2D\) functional. If \(0\ne q\in I_r\) with \(r\ge1\), choose two real forms \(F,G\in I_D\) such that \((q,F,G)\) is an Artinian complete intersection. First choose \(F\) avoiding the components of \(V(q)\), then choose \(G\) nonzero at the finitely many projective zeros of \(q,F\). These are choices outside finite unions of proper real linear subspaces. The Hilbert series of this complete intersection has top degree \(2D+r-3\). Since the original functional descends to its degree-\(2D\) piece, \(r\ge3\). When \(r=3\), the functional is nonzero on the one-dimensional socle and the perfect complementary-degree pairings give \(I=(q,F,G)\).

\subsection{2.1 Sharp entanglement threshold at arbitrary local rank}\label{s-2-1}

\begin{paperresult}{Theorem 1}{res-1}
For \(d\ge3\) and \(N\ge3\), the least operator rank of an entangled CS state of actual local rank \(N\) is \(3d+N-5\). Every entangled state \(\rho\) at that rank has a unique decomposition, up to normalization and ordering, into one positive ternary non-point extreme summand \(\tau\) of rank \(3d-2\) and \(N-3\) positive coherent projectors. The coherent directions are independent modulo the local support of \(\tau\). The minimal cone face is simplicial of dimension \(N-2\), and the two-party reduced state has rank \(N+3\). At the threshold rank, this face dimension characterizes entanglement. There is no one-party extension of \(\rho\) invariant under permutations of all \(d+1\) parties and PPT across a single-party cut.

\end{paperresult}

Write \(R_\rho\) for the operator rank and \(N_\rho\) for the one-party rank of a CS cone element \(\rho\). We work on its actual local support. For \(0\preceq \tau\preceq\rho\), with \(\tau\) also CS, one has \[
R_\rho-R_\tau\ge N_\rho-N_\tau. \tag{1}
\] To prove this, let \(W_\tau\subset W_\rho\) be the operator ranges and \(H_\tau\subset H_\rho\) the local supports. Contract \(d-1\) slots with a general real covector \(\ell\). On \(W_\rho\) this map surjects onto \(H_\rho\), since its dual is the injection \(\ell^{d-1}:A_1(\rho)\to A_d(\rho)\) from generic multiplication. Its restriction maps \(W_\tau\) into \(H_\tau\). Thus \(W_\rho/W_\tau\) surjects onto \(H_\rho/H_\tau\), giving (1).

For a non-point extreme ray of local rank \(n\ge4\), the two-step gap and the \hyperlink{small-hilbert-bound}{small Hilbert-value bound} proved above give \[
R_\tau\ge n+3(d-2)+2=3d+n-4. \tag{2}
\] For \(n=3\), Blekherman's bound \cite[Corollary 2.2]{cmp2} gives \(R_\tau\ge3d-2\) and its ternary examples attain equality. For \(n\le2\), all CS states are separable \cite[Theorem 18]{cmp1}. A ray on its actual local support is extreme there if and only if it is extreme in the larger cone, since every PSD summand has range contained in that of the ray and hence lies on the same local support. Consequently every non-point extreme ray satisfies \(R_\tau-N_\tau\ge3d-5\), with equality possible only at \(N_\tau=3\) and \(R_\tau=3d-2\).

The trace-one base of the CS cone is compact. A state therefore has a finite extreme-ray decomposition. An entangled state must have a non-point summand \(\tau\), so (1) yields \(R_\rho-N_\rho\ge3d-5\). For attainment, embed a ternary extreme ray of rank \(3d-2\) and add one positive coherent projector on each of \(N-3\) independent complementary axes. The ranks add and the projection onto the ternary support recovers the entangled ray. This proves the threshold in \hyperlink{res-1}{Theorem~1}. In particular all CS states below that rank are fully separable.

At equality every non-point summand has local support of dimension three and rank \(3d-2\). If two such supports \(H,H'\) differ and \(m=\dim(H\cap H')\), the two operator ranges intersect in a space of dimension at most \(\dim\operatorname{Sym}^d(H\cap H')\), because \(\operatorname{Sym}^d H\cap\operatorname{Sym}^d H'=\operatorname{Sym}^d(H\cap H')\). This identity follows by taking a basis adapted to the intersection. The sum \(\sigma\) of the two summands therefore satisfies \[
R_\sigma-N_\sigma\ \ge\
2(3d-2)-\dim\operatorname{Sym}^d(H\cap H')-(6-m).
\] For \(m=0,1\) the right side is \(6d-10\). For \(m=2\) it is \(5d-9\). Both exceed \(3d-5\) for \(d\ge3\), in contradiction to (1). All non-point summands thus have the same local support \(H\).

Put \(q=N-3\). Choose \(q\) coherent directions \(v_1,\ldots,v_q\) from the decomposition with independent classes modulo \(H\). Their powers are independent modulo \(\operatorname{Sym}^d H\). Together with the range \(W_\tau\) of any one ternary extreme summand they span the full operator range by the rank at equality. If a coherent power of \(w=h+\sum_i c_i v_i\) belongs to this range, the mixed tensor coordinates imply either \(w\in H\) or \(w\) belongs to one of the lines \(\mathbb Rv_i\). Indeed \(v_i^{d-1}v_j\) excludes two nonzero \(c_i\), and \(v_i^{d-1}h\) excludes a nonzero \(h\) when \(c_i\ne0\). The range \(W_\tau\) contains no coherent power, since \(I_d(\tau)\) has no projective base point. It follows that the only coherent directions in the full range are the \(q\) selected lines.

In coordinates adapted to \(H\oplus\bigoplus_i\mathbb Rv_i\), every CS element with range contained in the full range has only pure-block moments. A mixed monomial of degree \(2d\) admits a split into two degree-\(d\) monomials with at least one mixed, and the corresponding middle-matrix row or column vanishes. Each block is positive semidefinite. The \(H\) block has range in \(W_\tau\), so by extremality it is a nonnegative multiple of \(\tau\). Each line block is a nonnegative multiple of the corresponding coherent projector. These \(q+1\) rays are independent and generate the minimal face, proving the stated uniqueness and dimension. The \hyperlink{ternary-low-degree}{low-degree observation} in three variables gives \(I_2=0\) for the ternary core. Thus the two-party rank of the core is six and the other blocks add one apiece, giving \(N+3\). A separable state of rank \(R_\rho\) has at least \(R_\rho\) linearly independent coherent vectors in its range. The associated projectors are linearly independent, as follows by testing against the dual basis of these vectors. The minimal face of a separable state therefore has dimension at least \(R_\rho\). At rank \(3d+N-5\), this proves that the face dimension \(N-2\) characterizes entanglement.

\textbf{Corollary.} A non-point extreme CS state admits no positive CS extension by one party. The same holds for every entangled CS state of local rank \(N\) and operator rank \(3d+N-5\). First consider a non-point extreme \(\tau\). If \(L'\) were a positive extension of its functional \(L\), the slices \(L'(a^2\,\cdot)\) for real linear forms \(a\) would all lie in the minimal face of \(L\). For unit \(a\), complete \(a\) to an orthonormal basis and sum the slices to get \(L\). Scaling handles general \(a\). By polarization, \(L'(ab\,\cdot)=c(a,b)L\) for a real PSD bilinear form \(c\) with positive trace. Diagonalize \(c\) in an orthonormal basis with \(c(x_1,x_1)>0\). For \(j>1\) and \(p\in S_{2d-1}\), \[
c(x_1,x_1)L(x_jp)=L'(x_1^2x_jp)=c(x_1,x_j)L(x_1p)=0.
\] The local rank of \(L\) would be one, a contradiction. Now consider an arbitrary equality state and its direct sum of local supports just proved. Every quadratic product of linear coordinates from distinct blocks vanishes in its apolar ideal. An extension has the same two-party marginal and hence the same quadratic kernel, by the marginal kernel identification. Therefore it also has only pure-block moments and splits positively. Contraction for the partial trace is by the Euclidean quadratic \(g\) in the original coordinates. This operation sends a moment functional supported in one block to a moment functional on the same block, for any choice of the complementary subspaces. By uniqueness of the block decomposition, the partial trace of the ternary block equals the extreme core. This is impossible.

\textbf{Proposition (Permutation-invariant PPT extensions).} Let \(\rho\) be a CS state on \(d\ge2\) parties. A state \(\Omega\) on \(d+1\) copies of the same local space with marginal \(\rho\) is completely symmetric if and only if it is invariant under all \(d+1\) party permutations and PPT across one single-party cut.

For the forward implication, complete symmetry gives both properties directly. For the reverse implication, write \(\Gamma_1\) for partial transpose in the first party in the CS basis. By permutation invariance, we may take \(\Omega^{\Gamma_1}\succeq0\). Write \(P^-_{ij}\) for the orthogonal projector onto the subspace antisymmetric in parties \(i,j\). The two-party marginal on any pair is a two-party marginal of \(\rho\), so it is CS and is supported on the symmetric subspace of that pair. Positivity and \[
\operatorname{tr}(\Omega P^-_{ij})=0
\] therefore force \(\Omega\) to be supported on \(\operatorname{Sym}^{d+1}\), the common kernel of all \(P^-_{ij}\). For \(i,j\ne1\), the same trace equation holds for \(\Omega^{\Gamma_1}\), since the partial transpose acts trivially on \(P^-_{ij}\). For a pair \(1,j\), the corresponding marginal of \(\Omega^{\Gamma_1}\) is the partial transpose of the CS two-party marginal of \(\rho\), which equals that marginal. Thus the trace equation holds for this pair as well. Since \(\Omega^{\Gamma_1}\succeq0\), it too is supported on \(\operatorname{Sym}^{d+1}\). The PPT face characterization above now makes \(\Omega\) completely symmetric.

The preceding corollary therefore rules out these extensions for all non-point extreme CS states and all entangled CS states at the threshold of \hyperlink{res-1}{Theorem~1}. This gives the obstruction for the full-party version of the symmetric-extension question in \cite[Section VI]{cmp1}.

\subsubsection{Contraction and real local filtering}\label{contraction-and-real-local-filtering}

Let \(A\) be an invertible real matrix, let \(G=A^{\mathsf T}A\), and write \(g(x)=x^{\mathsf T}Gx\). For \(k\) parties, put \(\mathcal C_A^{(k)}(X)=A^{\otimes k}X(A^{\mathsf T})^{\otimes k}\). Define \(\operatorname{Tr}_{G}^{(s)}\) by contracting the last \(s\) parties against \(G^{\otimes s}\). Equivalently, filter the last \(s\) parties by \(A^{\otimes s}\) and then use the standard partial trace. The identity \[
\operatorname{Tr}_{s}\!\left(\mathcal C_A^{(d+s)}(\Omega)\right)
=\mathcal C_A^{(d)}\!\left(\operatorname{Tr}_{G}^{(s)}\Omega\right)
\] follows by taking the filter on the untraced parties outside the trace. The filters preserve complete symmetry and PPT by real tensor congruence. They send real product rays to real product rays and have inverses defined by \(A^{-1}\). In particular, \(\mathcal C_A^{(d)}\) identifies the \(g\)-weighted hierarchy on a projective support \(X\) with the standard partial-trace hierarchy on \(AX\). The equivalence holds on the cones, and density matrices are obtained by dividing by the positive trace. A common scalar multiple of \(A\) can always be chosen with operator norm at most one to realize the filter as a Kraus operator. That scalar cancels after normalization.

\Needspace{6\baselineskip}
\subsubsection{Sharp ranks at every finite PPT extension level}\label{sharp-ranks-at-every-finite-ppt-extension-level}

\begin{paperresult}{Theorem 1A}{res-1A}
Let \(d\ge3\), \(N\ge3\) and \(s\ge1\). Among entangled CS states of actual local rank \(N\) that admit a PPT symmetric extension by \(s\) copies of a single party, the least operator rank is \(3d+N-3\). It suffices to require PPT across one copied single-party cut. The minimum is the same at every finite level \(s\), and the intersection of all levels consists of the fully separable states.

\end{paperresult}

Write \(D=d+s\) with \(s\ge1\). Fix a cut between one party \(A\) and the remaining \(d-1\) parties \(B\). An extension on \(A_0,\ldots,A_s,B\) has marginal \(\rho\) on \(A_0B\) and is invariant under permutations of \(A_0,\ldots,A_s\). A PPT symmetric extension \cite{cmp7} is PPT across the bipartitions of these systems. On a CS marginal, this condition is equivalent to PPT across just \(A_0\).

To prove the equivalence, consider every pair consisting of a copy \(A_i\) and a party in \(B\). Its two-party marginal in the extension is the corresponding CS two-party marginal of \(\rho\), by permutation invariance of the copies. The same holds after partial transpose at \(A_0\), since the two-party marginals involving \(A_0\) are invariant under that transpose. If the extension and its partial transpose are positive, their supports therefore lie in the \(+1\) eigenspaces of all these pair swaps. The swaps form a connected graph on the \(D\) parties and generate all party permutations. Thus both operators have symmetric support. The PPT face argument of \hyperref[s-2]{Section~2} makes the extension completely symmetric. Conversely, a positive CS extension has all the required permutation and PPT properties.

On the polynomial side, this identifies the level-\(s\) extension test with existence of a functional \(L'\) on \(S_{2D}\) whose middle matrix is positive semidefinite and which satisfies \(L'(g^s f)=L(f)\) for all \(f\in S_{2d}\). Write \(\Sigma_{2D}\) for the cone of sums of squares from \(S_D\). The resulting cone of marginals is closed. On its trace-one slice it is a continuous image of the compact trace-one CS extension set, since \(L'(g^D)=L(g^d)\). Its dual is consequently \[
\{f\in S_{2d}:g^s f\in \Sigma_{2D}\}.
\] If \(L\) passes every level, then it is nonnegative on every strictly positive form of degree \(2d\): Reznick's positive-multiplier theorem \cite{cmp8} puts some \(g^s f\) in \(\Sigma_{2D}\). Approximating a nonnegative form \(f\) by \(f+\varepsilon g^d\) gives the same conclusion for nonnegative forms. Their dual cone is the cone generated by real point evaluations, whose trace-one base is the convex hull of the evaluations at unit vectors. Thus \(L\) is fully separable. Every fully separable CS state extends by adding copies to each coherent projector.

\textbf{Lower bound on the ranks of entangled marginals.} Let \(M\) be a non-point extreme functional of order \(2D\) and actual local rank \(n\). We claim that \[
h_j(M)\ \ge\ 3j+n-3\qquad (3\le j<D).
\] For \(n\ge4\), this is the consequence \(\delta_i\ge3\) for \(2\le i<D\) of the two-step rank gap and the \hyperlink{small-hilbert-bound}{small Hilbert-value bound}. For \(n=3\), non-point extremality implies entanglement across the one-party cut. The PPT rank criterion \cite{cmp4} gives \(h_D>h_{D-1}\), since \(h_{D-1}\ge3\) by generic multiplication. Strict middle growth then gives \(\delta_{D-1}\ge2\): a value of zero forces the next value to be zero, and a value of one gives a strict upper bound of one for \(\delta_D\).

If some \(\delta_k\le2\) with \(2\le k\le D-2\), the \hyperlink{small-hilbert-bound}{small Hilbert-value bound} and \(\delta_{D-1}\ge2\) force \(\delta_k=\delta_{k+1}=2\). Put \(J=(I_{\le k})\) and let \(\bar J\) be its image in the binary ring \(T=S/(\ell)\) for a general \(\ell\). Its degree-\(k\) subspace \(V=\bar J_k\) has dimension \(k-1\), and \(\bar J_{k+1}=T_1V\) because \(J\) is generated through degree \(k\). The quotient by \(\bar J\) surjects onto the apolar section, whose degree-\((k+1)\) dimension is two. Hence \(\dim T_1V\le k\). The \hyperlink{binary-minimal-growth}{binary minimal-growth argument} gives equality and \(V=q_0T_{k-2}\) for a quadratic \(q_0\). Every generator of \(\bar J\) is divisible by \(q_0\): multiply it by a suitable power of a linear form coprime to \(q_0\) to bring it into \(V\). Thus \(\bar J\subset(q_0)\) and they agree in degrees at least \(k\). The finite-length quotient \((q_0)/\bar J\) has regularity at most \(k-1\), and the regularity inequality for the exact sequence gives \(\operatorname{reg}\bar J\le k\).

Multiplication by \(\ell\) on \(S/J\) is injective out of degrees \(j\le k\): \(J_j=I_j\), \(J_{j+1}\subset I_{j+1}\), and the apolar multiplication is injective in this range. \hyperlink{section-lifting}{Lemma~2.1} makes \(J\) saturated and \(k\)-regular. Its Hilbert polynomial has first difference two, so its projective scheme is a plane curve of degree two. \hyperlink{associated-points}{Lemma~2.2} excludes zero-dimensional associated points. Hence \(J\) is height-one unmixed and, in the polynomial ring, is principal, generated by a quadratic \(q\). This gives \(q\in I_2\), contrary to the exclusion of quadratic apolar equations above.

It follows for \(n=3\) that \(\delta_k\ge3\) for \(2\le k\le D-2\). If \(\delta_{D-1}\ge3\) the claim follows directly. If \(\delta_{D-1}=2\), then \(D\ne4\) since \(I_2=0\) gives \(h_2=6\) and a drop to \(\delta_3=2\) would force \(\dim I_3=2\). Any one of those cubics would give a complete intersection of degrees \((3,4,4)\), with \(\dim I_3=1\). For \(D\ge5\), a cubic in \(I_3\) would give a complete intersection of degrees \((3,D,D)\) and hence \(\delta_{D-1}=3\), by the \hyperlink{ternary-low-degree}{low-degree observation}. Thus \(I_3=0\) and \(\delta_3=4\). In particular \(D\ge5\). The extra unit in degree three compensates for the final deficit and gives \(h_{D-1}\ge3(D-1)\). This proves the claim.

Decompose a positive CS extension into extreme rays. An entangled marginal requires at least one non-point extreme summand \(M\), because point evaluations remain positive point evaluations under contraction by \(g^s\). Its \(d\)-party marginal has local rank \(n\) and operator rank \(h_d(M)\), by the marginal kernel formula in \hyperref[s-1]{Section~1}. Since it is a positive summand of \(\rho\), inequality (1) gives \[
\operatorname{rank}\rho\ \ge\ h_d(M)+N-n\ \ge\ 3d+N-3.
\]

\textbf{Attainment at every finite level.} \hyperlink{res-1C}{Theorem~1C} constructs a rank-\(3d+N-3\) state that passes any prescribed level \(s\) and fails level \(s+1\). This part of \hyperlink{res-1C}{Theorem~1C} uses the lower bound above. Its block argument is independent of \hyperlink{res-1B}{Theorem~1B}. \hyperref[s-A]{Appendix~A} gives a separate proof of attainment by rational signed evaluations.

\textbf{Corollary (Extensions on several parties).} Prescribe finitely many extra copies of each party, with at least one extra copy in total. Require a joint extension invariant under permutations within each group of copies and PPT across every bipartition. The least rank of an entangled CS state admitting this extension is also \(3d+N-3\). Tracing out the extra copies of all but one extended party yields the lower bound from \hyperlink{res-1A}{Theorem~1A}. The construction at the total number of extra copies gives a CS extension, which satisfies every required symmetry and partial transpose condition.

\Needspace{6\baselineskip}
\subsubsection{Cubic rigidity at the extension threshold}\label{cubic-rigidity-at-the-extension-threshold}

\begin{paperresult}{Theorem 1B}{res-1B}
For \(d\ge3\) and \(N\ge3\), let an entangled CS state of actual local rank \(N\) and rank \(3d+N-3\) admit a PPT symmetric extension by at least one copy. It has a unique decomposition into an entangled ternary core of rank \(3d\) and \(N-3\) positive coherent projectors on directions independent modulo its local support. The core's apolar ideal through degree \(d\) is generated by a reduced real plane cubic; each geometric component has dense real points. The quadratic marginal kernel recovers the ternary plane and the complementary points, and the cubic marginal kernel recovers the curve. The minimal cone face is the product of the Hankel cone on this cubic and \(N-3\) one-dimensional cones, of total dimension \(6d+N-3\). Every CS extension splits into the same blocks. Within its ternary block, every non-point extreme summand of an extension to \(D>d\) parties has ideal \((q,F,G)\), where \(q\) is the recovered cubic and \(F,G\) have degree \(D\). It therefore has rank \(3D-2\).

\end{paperresult}

\textbf{Lemma 2.3 (Multiplication with a three-dimensional target).} Let \(\mu:V\times V\to W\) be a symmetric real bilinear map with \(\dim V\ge3\) and \(\dim W=3\). Suppose its products span \(W\) and its common annihilator \(\{u:\mu(u,V)=0\}\) is zero. Then some multiplication map \(m_u=\mu(u,\cdot)\) is surjective.

Let \(r\) be the maximum rank and choose \(u\) at that rank with \(\mu(u,u)\ne0\). Such a choice exists since maximal rank and nonzero square are nonempty polynomial open conditions, the latter by polarization. Suppose \(r<3\), and put \(K=\ker m_u\) and \(W_0=m_u(V)\). For every \(v\in V\), the rank bound on \(m_{u+tv}\) implies \(m_v(K)\subset W_0\). Indeed, take \(r\) columns with independent images under \(m_u\) and append the image of any \(k\in K\). The coefficient of \(t\) in the resulting exterior product is the wedge of the \(r\) independent images with \(m_v(k)\), and must vanish. By symmetry the pairing modulo \(W_0\) therefore descends to \(V/K\) in both variables. If \(r=1\), the products on that one-dimensional quotient cannot span the two-dimensional \(W/W_0\). Thus \(r=2\).

Choose a scalar identification of \(W/W_0\). The induced scalar pairing on \(V/K\) is nonzero and has \(u+K\) in its radical. It therefore has rank one and is a nonzero scalar times \(\lambda\otimes\lambda\) for a linear form \(\lambda\) on \(V\) vanishing on \(K\) and \(u\). Put \(H=\ker\lambda\). For any \(v\) with \(\lambda(v)\ne0\), the projection of \(m_v(V)\) onto \(W/W_0\) is nonzero, so \(m_v|_H:H\to W_0\) has rank at most one. Its two-by-two minors vanish for all \(v\) by polynomial continuation.

A linear space of rank-at-most-one maps has either a common image line or a common kernel hyperplane. To see this, write each nonzero map as a vector times a covector. If two have independent image vectors, their covectors must be proportional, since their sum still has rank at most one. Comparing any other nonzero map with one of this pair forces its covector to be proportional to theirs. Applied to the maps \(m_v|_H\), a common image line is impossible because \(u\in H\) and the products \(m_v(u)=m_u(v)\) span \(W_0\). A common kernel hyperplane would give a nonzero common annihilator because \(\dim H=\dim V-1\ge2\). This proves surjectivity.

\Needspace{5\baselineskip}
\hypertarget{higher-local-rank}{}\textbf{Lemma (Higher local rank).} Let \(I\) be the apolar ideal of a non-point extreme functional of degree \(2D\), with \(D\ge4\) and actual local rank \(n=h_1\ge4\). Then \(\delta_2\ge4\), and \(h_j\ge3j+n-2\) for \(2\le j<D\).

\textbf{Proof.} The proof of \hyperlink{res-1A}{Theorem~1A} gives \(\delta_j\ge3\) for \(2\le j<D\). It suffices to rule out \(\delta_2=3\).

Suppose \(\delta_2=3\) and put \(B=S/((I_2),\ell)\) for a general real linear form \(\ell\). Here \((I_2)\) denotes the ideal generated by \(I_2\). Then \(\dim B_2=3\), because \((I_2)\) and \(I\) agree through degree two. The quadratic section argument in \hyperref[s-2-2]{Section~2.2} gives zero common annihilator for \(B_1\times B_1\to B_2\). Since \(B\) is standard graded, products span \(B_2\). Lemma 2.3 gives \(v\in B_1\) with \(B_2=vB_1\). Therefore \[
B_3=B_1B_2=vB_2 .
\] In particular \(\dim B_3\le3\), while its surjection onto the degree-three apolar section gives \(\dim B_3\ge\delta_3\ge3\). Thus multiplication by \(v\) from \(B_2\) to \(B_3\) is an isomorphism.

If \(w\in B_1\) is killed by \(v\), then \(wB_1\) is in the kernel of this second map, and therefore \(w=0\). It follows that \(n=4\). The quotient by \(v\) has Hilbert series \(1+2t\) and defining ideal generated by all quadrics in two variables. Its defining ideal is \(2\)-regular. Apply Lemma 2.1 with \(k=2\) first to \(v\) on \(B\), using the injective maps out of degrees one and two. Then apply it to \(\ell\) on \(S/(I_2)\), since \((I_2)\) agrees with \(I\) through degree two and the apolar multiplication is injective into degree three. The two regular linear forms show that \(S/(I_2)\) is the Cohen--Macaulay coordinate ring of a degree-three curve in \(\mathbb P^3\), with Hilbert series \((1+2t)/(1-t)^2\).

Write \(R=S/(I_2)\). The degree-\(D\) apolar system \(I_D\) has no complex projective base point by extremality. Choose real \(F,G\in I_D\) such that their images in \(R_D\) have no common zero on \(\operatorname{Proj}(R_{\mathbb C})\). Indeed, choose \(F\) avoiding the finitely many conditions of vanishing identically on a component, then choose \(G\) nonzero at every point of the finite zero set of \(F\) on that curve. Each condition excludes a proper real linear subspace of \(I_D\). Their images in \(R\) form a homogeneous system of parameters. Since \(R\) is Cohen--Macaulay, they are a regular sequence. One can check regularity after localization at the homogeneous maximal ideal, since any nonzero graded kernel remains nonzero there. Thus \[
H_{R/(F,G)}(t)
= (1-t^D)^2 H_R(t)
= (1+2t)(1+t+\cdots+t^{D-1})^2 .
\] The polynomial on the right has degree \(2D-1\), so \(R_{2D}=F R_D+G R_D\). The original functional descends to \(R_{2D}\) because \((I_2)\subset I\). It annihilates both terms on the right because \(F,G\in I_D\). This contradicts its nonzero value on some form of degree \(2D\), proving \(\delta_2\ge4\).

\Needspace{5\baselineskip}
\hypertarget{ternary-equality}{}\textbf{Lemma (Ternary equality).} Let \(I\) be the apolar ideal of a non-point extreme functional of degree \(2D\) and actual local rank three. Suppose \(h_d=3d\) for some \(3\le d<D\). Then \(I=(q,F,G)\) is a complete intersection with degrees \((3,D,D)\), where \(q\) defines a reduced plane cubic with Zariski-dense real points in every geometric component.

\textbf{Proof.} The proof of \hyperlink{res-1A}{Theorem~1A} gives \(\delta_j\ge3\) for \(2\le j\le D-2\) and \(\delta_{D-1}\ge2\). If \(d\le D-2\) or \(\delta_{D-1}\ge3\), equality gives \(\delta_2=\cdots=\delta_d=3\).

The only additional possibility is \(d=D-1\) and \(\delta_{D-1}=2\). The preceding lower-bound proof gives \(D\ge5\), \(I_3=0\), and \(\delta_3=4\). Since \(h_{D-1}=3(D-1)\), the increments with \(2\le j\le D-2\) with \(j\ne3\) are all three, so \(\delta_4\le3\) also when \(D=5\). Thus \(h_4\le13\) and there is a nonzero quartic \(Q\in I_4\). The \hyperlink{small-hilbert-bound}{small Hilbert-value bound} gives \(\delta_D=1\), hence \(h_D=3D-2\).

Use the complete-intersection construction above with \(Q\) to obtain \(C=(Q,F,G)\subset I\) of degrees \((4,D,D)\). Its socle degree is \(2D+1\). The perfect pairing of \(S/C\) between degrees one and \(2D\) writes the functional defining \(I\) as \(p\mapsto\lambda(ap)\), where \(\lambda\) is a nonzero socle functional and \(a\) is linear. The complementary-degree pairings then give \(I=C:a\). The image of \(Q\) in \(S/(a)\) is nonzero, because a factorization \(Q=aQ'\) would put \(Q'\in I_3\).

In the binary ring \(S/(a)\) the degree-\((D+1)\) quotient by \(Q\bmod a\) has dimension four. Since the images of \(Q,F,G\) have no common projective zero, a real linear combination of the images of \(F,G\) is a nonzero divisor modulo the image of \(Q\). Its two independent linear multiples give \[
\dim(S/(C,a))_{D+1}\le2.
\] In three variables the Hilbert series of the complete intersection gives \(\dim(S/C)_{D+1}=\binom{D+3}{2}-\binom{D-1}{2}-2\binom{3}{2}=4D-4\). The image of multiplication by \(a\) from degree \(D\) to \(D+1\) has dimension \(h_D\), since its kernel is \((C:a)_D/C_D\). Therefore \[
h_D\ge (4D-4)-2=4D-6>3D-2
\qquad (D\ge5).
\] This contradicts the exceptional case and also gives a rank bound for any ternary apolar ideal with base-point-free \(I_D\), \(I_3=0\), and \(I_4\ne0\).

We now have \(I_2=0\) and \(\dim I_3=1\) on the ternary support. Let \(q\) generate \(I_3\). A repeated geometric factor of a real cubic would be a real linear factor, so \(q=\ell^2m\) for real linears. Then \((\ell m)^2=mq\in I\), contrary to \(I_2=0\). A conjugate pair of different complex lines, or a real conic with no smooth real point as a geometric component, would give \(q=mQ_0\) with \(Q_0\) a nonzero sum of real linear squares up to sign. The sum of squares \(m^2 Q_0\) in \(I\) again gives a nonzero quadratic in \(I\), by positivity of the pairing on degree two. A geometrically irreducible real cubic has a smooth real point, since a general real line meets it in three distinct smooth complex points. Thus the cubic is reduced, and real points are dense in each geometric component. A plane cubic is connected and has arithmetic genus one.

The cubic case of the \hyperlink{ternary-low-degree}{low-degree observation} gives \(I=(q,F,G)\) with \(F,G\) of degree \(D\). Its Hilbert series yields \(h_D=3D-2\), and below degree \(D\) the ideal consists of the multiples of \(q\). This identifies the kernel of every proper marginal of the summand.

\textbf{Proof of Theorem 1B.} Write \(\rho\) as in \hyperlink{res-1B}{Theorem~1B} and take any of its positive CS extensions to \(D>d\) parties. In an extreme-ray decomposition of the extension there is at least one non-point extreme summand. For each such summand, projected to \(d\) parties, the rank inequalities in \hyperlink{res-1A}{Theorem~1A} must both be equalities. The \hyperlink{higher-local-rank}{higher-local-rank lemma} forces local rank three. The \hyperlink{ternary-equality}{ternary equality lemma} then identifies its ideal. Its marginal has range of dimension \(3d\) in the \(d\)-fold symmetric power of its local support, and that range is the annihilator of the degree-\(d\) multiples of the recovered cubic.

All non-point summands have the same local support \(H\). For two distinct ternary supports \(H_1,H_2\), write \(m=\dim(H_1\cap H_2)\le2\). Extend a basis of \(H_1\cap H_2\) first in each space and then in their sum. The associated monomial basis shows that \(\operatorname{Sym}^d(H_1)\) and \(\operatorname{Sym}^d(H_2)\) intersect in \(\operatorname{Sym}^d(H_1\cap H_2)\). This space has dimension at most \(d+1\). The sum of the two positive marginals has local rank \(6-m\) and operator rank at least \(6d-(d+1)\). Its operator rank minus local rank is thus at least \(5d-7>3d-3\) for \(d\ge3\), contrary to (1). Once the support is fixed, their ranges coincide as well: otherwise the sum would have rank greater than \(3d\) on local support \(H\), again contradicting (1). Their cubic ideals thus coincide.

Choose \(N-3\) of the coherent directions whose classes span the complement modulo \(H\). Their powers together with the common ternary range fill the range of \(\rho\) by the specified rank. In coordinates adapted to this direct sum, any coherent power in the range is either in the ternary range or belongs to one of the selected lines, by inspection of its mixed coordinates. Every positive CS functional with range in that of \(\rho\) splits into these blocks, since all mixed top moments vanish, as in \hyperref[s-2-1]{Section~2.1}. The ternary block of the face is exactly the positive Hankel cone on the cubic: its kernel contains the degree-\(d\) multiples of \(q\), which multiply into all degree-\(2d\) multiples of \(q\). The core of \(\rho\) has full middle rank on the cubic, so this cone has dimension \(\dim(S_H/(q))_{2d}=6d\), where \(S_H\) is the polynomial ring on \(H\). Each coherent line adds one to the face dimension. The core is entangled, since the other blocks are separable.

For \(1\le j\le d\), the apolar kernel of the ternary block in degree \(j\) equals the degree-\(j\) part of \((q)\). Indeed, if \(p\) is in this kernel, then \(p\lambda^{d-j}\) is a multiple of \(q\) for any real linear form \(\lambda\) coprime to \(q\), which implies that \(q\) divides \(p\). Together with the mixed quadrics, this gives \(h_j(\rho)=3j+N-3\). The quadratic kernel defines the disjoint union of the ternary plane and the \(N-3\) complementary points in projective space. Adding the cubic kernel cuts the plane down to the cubic. These kernels determine the blocks, and the block restrictions give the unique decomposition.

The two-party marginal kernel is the same for \(\rho\) and any CS extension. Therefore the cross-block quadrics also vanish in every extension, and its positive blocks restrict under contraction to the corresponding blocks of \(\rho\). The cubic kernel is likewise shared, so the ternary block of every extension is supported on the same cubic. Its non-point extreme summands have the complete intersections described above. This proves \hyperlink{res-1B}{Theorem~1B}.

In particular, an entangled \(\rho\) at the first extension threshold must have marginal ranks \(h_j=N+3(j-1)\) for all \(1\le j\le d\) to pass even one round. At rank \(3d+N-3\), a two-party marginal rank different from \(N+3\) therefore rules out the first PPT extension of an entangled state.

For an exact example with complementary points, use the \(s=2\) row of the rational three-qutrit construction in \hyperref[s-A]{Appendix~A} and add evaluations of weight one on the two complementary coordinate axes in local rank five. In the ternary coordinates \((y_0,y_1,y_2)=(x_0+x_1+x_2,tx_0,tx_1)\), the cubic is \(y_1y_2(ty_0-y_1-y_2)\). Direct rational evaluation gives marginal ranks \((h_1,h_2,h_3)=(5,8,11)\) and puts that cubic in the three-party kernel. The mixed quadrics and this cubic recover the triangle and the two complementary points.

\hypertarget{extension-ranks}{}\textbf{Corollary 2.5 (Ranks of extensions at the threshold).} Let \(\rho\) satisfy the hypotheses of \hyperlink{res-1B}{Theorem~1B}, and let \(\Omega\) be a PPT symmetric extension to \(D>d\) parties. Then \[
\operatorname{rank}\Omega\in\{3D+N-5,\ 3D+N-4,\ 3D+N-3\}.
\] The smallest rank occurs exactly when the ternary block of \(\Omega\) spans a non-point extreme ray. Moreover, fix a reduced real plane cubic \(X\) with Zariski-dense real points in each geometric component, and a positive definite contraction quadratic \(g\) on its plane. If level \(D-d\) on \(X\) is inexact, then each of the three ranks is attained by a \(D\)-party extension of an entangled rank-\(3d+N-3\) marginal with that ternary cubic and with actual local rank \(N\). The marginal may vary with the extension rank.

\textbf{Proof.} By \hyperlink{res-1B}{Theorem~1B} the line blocks contribute \(N-3\) to the rank, and the ternary block is positive on the degree-\(D\) quotient by the cubic, of dimension \(3D\). An extreme-ray decomposition of that block must contain a non-point term, because its contraction is entangled. Each such term has rank \(3D-2\). If the block itself has rank \(3D-2\), it has the same range as any such term and hence lies in that term's one-dimensional minimal face. The converse follows from the rank of a non-point extreme term.

For attainment, write \(R\) for the coordinate ring of \(X\). By \hyperlink{support-prop}{Proposition~3.1}, inexactness gives a positive functional on \(R_{2D}\) with negative value on \(g^{D-d}f\) for some \(f\in P_{2d}(X)\). The cone of all positive functionals on \(R_{2D}\) is a face of the ambient CS cone, cut out by vanishing on the degree-\(D\) multiples of the cubic in the middle kernel. Decompose the positive functional into ambient extreme rays in this face. One term \(M\) still has negative value on \(g^{D-d}f\). It is a non-point ray, since this form is nonnegative at every real point of \(X\). It has local rank three because all binary CS states are separable. \hyperlink{res-1A}{Theorem~1A} gives marginal rank at least \(3d\), and support on \(X\) gives rank at most \(3d\). The ternary equality case in \hyperlink{res-1B}{Theorem~1B} now gives \(\operatorname{rank}M=3D-2\).

Let \(K\) be the two-dimensional kernel of the middle pairing of \(M\) on \(R_D\). Real point evaluations on \(X\) span \(K^*\) by real density. Choose two whose restrictions to \(K\) are independent. Adding a small positive multiple of one evaluation, or of both, cuts \(K\) by one or by two independent linear equations, raising the rank by one or two. Negativity on \(g^{D-d}f\) persists for small coefficients and the contracted rank remains \(3d\), since the original contraction was already full rank on \(R_d\). The three ternary ranks are thus attained. For \(N>3\), adjoin \(N-3\) positive coherent projectors on complementary axes orthogonal for the contraction metric and contract. Ranks add, and projection onto the ternary block preserves the entanglement certificate. Finally normalize each extension by its trace. This proves the corollary.

\subsection{2.2 A constraint on the quadratic section}\label{s-2-2}

\textbf{Common annihilator on the quadratic section.} Let \(L\) span a non-point extreme ray of the CS cone in degree \(2d\), with \(d\ge3\), and work on its actual local support, so \(I_1=0\). Choose a real linear form \(\ell\) whose multiplications \(A_1\to A_2\) and \(A_2\to A_3\) are injective, as holds for a general \(\ell\) by generic multiplication. Put \(R=S/(I_2)\), where \((I_2)\) denotes the ideal generated by \(I_2\), and \(B=R/(\ell)\). Then \(B_1\) has no nonzero element annihilating \(B_1\).

Indeed, a lift \(p\) of such an element defines \(T:A_1\to A_1\) by \(px=\ell Tx\) in \(A_2\). The first injectivity makes \(T\) well defined. For the positive definite inner product \(L(\ell^2xyg^{d-2})\), \(T\) is self-adjoint, since moving \(T\) from either argument gives \(L(\ell pxyg^{d-2})\). If \(Tx=\lambda x\) and \(Ty=\mu y\) with \(\lambda\ne\mu\), then computing \(pxy\) in both ways gives \((\lambda-\mu)\ell xy=0\) in \(A_3\). The second injectivity gives \(xy=0\) in \(A_2\). All mixed top moments between distinct eigenspaces therefore vanish, and \(L\) is the sum of its positive restrictions to the eigenspaces. Each eigenspace contributes a nonzero restriction, since its degree-one vectors belong to the actual support. Extremality forces one eigenvalue, and \(T\ell=p\) shows that \(p\) vanishes modulo \(\ell\).

\subsection{2.3 Ambient witnesses and open separation of levels}\label{s-2-3}

\textbf{Proposition 2.4.} Let \(q\) be a nonzero real form of degree \(r\le d\) and let \(g\) be positive definite. Suppose \(L(g^d)=1\) and the state functional \(L\) annihilates \((q)_{2d}\) and satisfies \(L(f)<0\) for some \(f\in S_{2d}\) with \(g^s f\) a sum of squares modulo \((q)\). Then there are \(\epsilon,C>0\) such that \[
F=f+\epsilon g^d+C q^2 g^{d-r},
\qquad L(F)<0,\qquad g^s F\in\Sigma_{2(d+s)}.
\] Thus \(F\) excludes an extension at level \(s\) in the ambient state cone.

Put \(D=d+s\) and choose a basis of \(S_D\) consisting of a complement of \(qS_{D-r}\) and the forms in \(q\) times a basis of \(S_{D-r}\). Lift the squared forms modulo \((q)\) to the complement. The resulting Gram matrix has positive semidefinite upper-left block. The remaining difference from \(g^s f\) belongs to \((q)_{2D}\). Every term of this difference can be represented using a product with at least one factor in \(qS_{D-r}\), since \(S_D S_{D-r}=S_{2D-r}\). Adding those entries leaves the upper-left block positive semidefinite. Multiplication of \(\epsilon g^d\) by \(g^s\) adds a positive definite Gram matrix, by expanding a positive sum of squares of independent linear forms to the power \(D\). The term \(C q^2 g^{D-r}\) adds a positive definite matrix times \(C\) on the lower-right block. The Schur complement is positive for sufficiently large \(C\). Finally \(L(F)=L(f)+\epsilon<0\) for \(0<\epsilon<-L(f)\). This proves the proposition.

\textbf{Consecutive levels separate on an open set.} Suppose a trace-one CS state passes level \(s\) and fails level \(s+1\). The level-\(s+1\) cone is closed by \hyperref[s-2-1]{Section~2.1}. Thus the original state remains outside it under sufficiently small perturbations. Mix a positive fraction of the separable state \(\sigma_d\) obtained by averaging real pure product states \(x^{\otimes d}\) over the unit sphere of the actual local support. Its natural \((d+s)\)-party extension is positive definite on the symmetric subspace, since the integral of the square of any nonzero homogeneous form there is positive. The mixture therefore has a positive definite extension. The contraction map on the vector spaces of completely symmetric tensors is surjective, since its dual is multiplication by the nonzero polynomial \(g^s\). It follows that the mixture is interior to level \(s\) in the trace-one affine space of CS tensors. Its operator rank on the symmetric subspace is full. \hyperlink{res-1C}{Theorem~1C} thus yields open full-rank regions separating any two consecutive positive levels for every \(d,N\ge3\).

For the triangle example of Section~1.1, \hyperref[s-A-3]{Appendix~A.3} gives a rational ambient witness \(F\) with \[
L_\rho(F)=-6/T,\qquad
L_{\sigma_3}(F)=\frac{2113141}{286720}.
\] For \(0<\eta<46448640/17824696077844693\), the full-rank state \((1-\eta)\rho+\eta\sigma_3\) passes the first level and fails the second. The rational endpoint, about \(2.61\cdot10^{-9}\), is a certified lower bound for the noise threshold of this example.

\Needspace{20\baselineskip}
\section{3. Finite exactness and extension depth on curves}\label{s-3}

The table collects the tests for every finite case of \hyperlink{res-1D}{Theorem~1D}. A split node is an ordinary double point with two distinct real tangent directions.

\begin{longtable}[]{@{}
  >{\raggedright\arraybackslash}p{(\columnwidth - 2\tabcolsep) * \real{0.5000}}
  >{\raggedright\arraybackslash}p{(\columnwidth - 2\tabcolsep) * \real{0.5000}}@{}}
\toprule\noalign{}
\begin{minipage}[b]{\linewidth}\raggedright
Cubic type
\end{minipage} & \begin{minipage}[b]{\linewidth}\raggedright
First exact level or test for level \(s\)
\end{minipage} \\
\midrule\noalign{}
\endhead
\bottomrule\noalign{}
\endlastfoot
Smooth & One Gram matrix of order \(3s\). At level one the matrix is explicit. \\
Geometrically irreducible with a split real node & One binary certificate with a semidefinite lift of order \(3s+2\). \\
Line and smooth conic meeting at two real transverse points & One Gram matrix of order \(2s+1\). At level one the matrix is explicit. \\
Line and smooth conic meeting at conjugate points in \(V(g)\) & Level one. \\
Three nonconcurrent real lines & The ceiling of \(\pi\) divided by the sum of the three projective angles. \\
\end{longtable}

\Needspace{6\baselineskip}
\subsection{Restriction to an algebraic support}\label{s-3-support}

\hypertarget{support-prop}{}\textbf{Proposition 3.1.} Let \(X\) be a reduced real projective algebraic set with dense real points in every geometric component. Write \(J=I(X)\), \(R=S/J\), and assume \(J\) is generated in degrees at most \(d\). Let \(g\) be positive definite, let \(s\ge0\), and put \(D=d+s\). Define \(P_{2d}(X)\) to be the cone of forms in \(R_{2d}\) nonnegative on \(X(\mathbb R)\), and \(\Sigma_{2D}(X)\) the sums of squares of forms in \(R_D\). At level zero the extension cone is the CS cone on this support. At each level, including zero, the ambient level-\(s\) states annihilating \(J_{2d}\) are precisely the contractions of positive functionals on \(R_{2D}\) by \(g^s\). On that support, the level-\(s\) cone equals the separable cone if and only if \[
g^s P_{2d}(X)\subseteq\Sigma_{2D}(X).
\] When inclusion fails, an entangled level-\(s\) state can be chosen with marginal ranks \(\dim R_j\) for \(1\le j\le d\).

\textbf{Proof.} Work first with unnormalized state cones. If \(M\) is an ambient CS extension of \(L\) and \(p\in J_a\) is a generator of degree \(a\le d\), then \[
M\!\left(p^2g^{D-a}\right)=L\!\left(p^2g^{d-a}\right)=0 .
\] By the marginal-kernel identity, \(p\in\operatorname{Ann}(M)_a\). That identity also holds for positive definite \(g\) after a real change of linear coordinates. The apolar annihilator is an ideal, so \(M\) vanishes on \(J_{2D}\). Conversely a positive functional on the quotient pulls back to an ambient CS extension, giving the claimed identification and the closedness inherited from the ambient cone. If \(L\) on the support is separable, write it as a positive sum of real evaluations. The same nonnegative expression \(p^2g^{d-a}\) forces every point in the sum to satisfy every generator. Hence the separable cone on the support is the closed cone spanned by real evaluations on \(X\), with compact trace-one base.

The cone \(\Sigma_{2D}(X)\) is closed. Choose finitely many real evaluations spanning \(R_D^*\) using the density assumption. Their sum in degree \(2D\) induces a positive definite form on \(R_D\) and bounds the traces of Gram matrices representing a convergent sequence in \(\Sigma_{2D}(X)\). Dualizing contraction now shows that the dual of the level-\(s\) cone is the inverse image of \(\Sigma_{2D}(X)\) under multiplication by \(g^s\). Since \(g\) is positive on real representatives, that inverse image lies in \(P_{2d}(X)\), which is the dual of the separable cone. This gives the criterion.

If the criterion fails for \(f\in P_{2d}(X)\), separate \(g^s f\) from the closed SOS cone. Add a small multiple of the positive definite sum of evaluations above to the separating functional \(M\) so that \(M\) is positive definite and \(M(g^s f)<0\). Choose a real linear form \(\ell\) nonzero on every component. The quadratic \(g\) can be written as a sum of squares including a positive multiple of \(\ell^2\). Thus the \(j\)-party moment form \(p\mapsto M(g^{D-j}p^2)\) is positive on nonzero \(p\in R_j\), since \(\ell^{D-j}p\ne0\). Normalize by its positive trace. The resulting state is entangled by its negative value on \(f\). This proves the rank assertion.

\subsection{Endpoint correlations and line arrangements}\label{endpoint-correlations-and-line-arrangements}

\begin{paperresult}{Theorem 1C}{res-1C}
Let \(G\) be a finite simple graph on \(n\) independent real unit vectors \(u_i\), with no isolated vertices, and let \(X\) be the union of the projective lines on its edges. Put \(\phi_{ij}=\arccos|\langle u_i,u_j\rangle|\) and define \[
J_0(\phi)=\{1\},\qquad J_1(\phi)=\{\cos\phi\},\qquad
J_s(\phi)=[\cos(\min(s\phi,\pi)),1]\quad(s\ge2).
\] For every \(d\ge3\), the level-\(s\) PPT extension cone on the degree-\(2d\) Hankel cone of \(X\) equals the separable cone if and only if for every choice of edge signs \(\eta_{ij}\in\{1,-1\}\) there is a real correlation matrix \(B\) with \(\eta_{ij} B_{ij}\in J_s(\phi_{ij})\). It suffices to vary signs on edges outside a spanning forest. In particular, every arrangement has a finite first exact level. If \(G\) has no \(K_4\) minor, that level is \[
s_0=\left\lceil\frac{\pi}{\ell(G)}\right\rceil ,
\] where \(\ell(G)\) is the least total angular length of a cycle, and \(\ell(G)=\infty\) for a forest. At any level where the criterion fails, there are entangled level-\(s\) marginals of local rank \(n\) and operator rank \(n+(d-1)|E(G)|\). In particular every pair of consecutive positive levels is separated by a CS state of rank \(3d+N-3\) and local rank \(N\), for all \(d,N\ge3\).

\end{paperresult}

Use the independent vertices \(u_i\) as a basis of their span and write \(R=\mathbb R[X]\). In positive degree, forms in \(R\) are binary forms on the edges agreeing at the vertices. The pure vertex monomials realize the common values, and the mixed monomials supported on a single edge realize the other coefficients. The ideal of \(X\) is generated by square-free nonedge quadrics and square-free cubics. In particular \(\dim R_j=n+(j-1)|E(G)|\) for \(j\ge1\). Write \(P_{2d}(X)\) for the forms nonnegative on \(X(\mathbb R)\) and \(\Sigma_{2j}(X)\) for sums of squares from \(R_j\). In general SOS multiplier bounds on projective curves allow the multiplier to depend on the input \cite{cmp10}. Here the partial trace fixes \(g(x)=\|x\|^2\).

\hypertarget{edge-endpoint-bound}{}\textbf{The two endpoint values on an edge.} Fix two unit vectors \(u,v\) in a Euclidean plane with angle \(\phi\in(0,\pi/2]\). For a real vector \(Q\) of degree-\(s\) binary forms with \(\|Q\|^2=g^s\), let \(c=\langle Q(u),Q(v)\rangle\). The set of all \(c\) is exactly \(J_s(\phi)\). For \(s=0\) the vector is constant. For \(s=1\) it is the restriction of a linear isometry and preserves the inner product. The set of \(c\) is convex by taking direct sums of two representations multiplied by square roots of convex weights.

Suppose \(s\ge2\). On the unit circle, \(Q\) is a unit vector-valued trigonometric polynomial of degree at most \(s\). At each point the scalar projection in the direction of its derivative has supremum norm at most one. Bernstein's inequality \cite{cmp9} therefore bounds the norm of that derivative by \(s\). The spherical distance between \(Q(u)\) and \(Q(v)\) is at most \(s\phi\), which proves \(c\ge\cos(\min(s\phi,\pi))\). If \(s\phi\le\pi\), the real and imaginary parts of \((x+iy)^s\) attain this bound in orthonormal plane coordinates.

For \(s\phi\ge\pi\), it suffices to find a homogeneous real scalar form \(F\) of degree \(s\) with \(|F|\le1\) on the unit circle and opposite values in \(\{1,-1\}\) at the two endpoints. The residual binary form \(g^s-F^2\) is nonnegative and hence is a sum of two real squares, by grouping conjugate factors and halving the even multiplicities at real roots. The two residual summands vanish at both endpoints, so adjoining them to \(F\) gives \(c=-1\). For even \(s\), choose a unit coordinate \(y\) with endpoint values \(0,\sin\phi\). Use the homogenization of the Chebyshev polynomial \(T_s(k y)\) on the circle, where \(k=\sin(\pi/s)/\sin\phi\le1\). The values at \(0\) and \(\sin(\pi/s)\) are opposite and have absolute value one. For odd \(s\), choose \(y\) perpendicular to the bisector, with endpoint values \(\pm\sin(\phi/2)\). Take \(k=\sin(\pi/(2s))/\sin(\phi/2)\le1\). Then \(T_s\) takes opposite unit values at \(\pm\sin(\pi/(2s))\). In both cases the homogeneity is obtained by replacing a term of degree \(j\) by that term times \(g^{(s-j)/2}\), using the parity of \(T_s\). Its boundedness on the circle follows from \(|T_s|\le1\) on \([-1,1]\).

For even \(s\), the upper endpoint \(c=1\) is attained by \(Q=g^{s/2}\). For odd \(s\ge3\), let \(x\) be the unit bisector coordinate and \(t=\cos(\phi/2)\). Use the homogeneous scalar form \(F=g^{(s-3)/2}(3xg/t-x^3/t^3)/2\). At both endpoints its value is one. On the circle, with \(z=x/t\), the identity \[
1-\left(\frac{3z-z^3}{2}\right)^2
=\frac{(1-z^2)^2(4-z^2)}4
\] gives \(|F|\le1\), since \(|z|\le\sqrt 2\). Complete \(F\) by a binary sum of squares as above. This attains the upper endpoint and proves the interval formula. In fixed vertex representatives the endpoint correlation interval on edge \(ij\) is \(\sigma_{ij}^{\,s}J_s(\phi_{ij})\), where \(\sigma_{ij}\in\{1,-1\}\) makes the angle between \(u_i\) and \(\sigma_{ij}u_j\) the projective angle.

\textbf{Which edge data must be glued.} The cone \(P_{2d}(X)\) is closed, pointed, and has a compact base. In fact, finitely many real evaluations span \(R_{2d}^*\), and their sum is strictly positive on every nonzero element of \(P_{2d}(X)\). Its unit slice is then compact. It is enough to test the inclusion \(g^sP_{2d}(X)\subseteq\Sigma_{2(d+s)}(X)\) on extreme rays of \(P_{2d}(X)\). On an edge where an extreme form \(f\) is nonzero, write it as \(h^2q\) with all its real zeros in \(h\) and with \(q\) strictly positive. If \(\deg q\ge2\), there is a nonzero binary form \(r\) of that degree vanishing at both endpoints. Replacing this edge by \(h^2(q\pm\varepsilon r)\) preserves nonnegativity for small \(\varepsilon>0\), and preserves all vertex values. These two forms are not proportional to \(f\), contrary to extremality. Consequently each nonzero edge restriction of an extreme ray is \(p_{ij}^2\), where all roots of \(p_{ij}\) are real, with multiplicities allowed.

For necessity of the correlation test, choose all edge squares \(p_{ij}^2\) with distinct real roots away from the vertices and endpoint values \(p_{ij}(u_i)=1\), \(p_{ij}(u_j)=\varepsilon_{ij}\in\{1,-1\}\). Such forms are obtained by prescribing the product of their roots in the vertex basis. They glue because the squared values are one. Any SOS representation has a vector of summands \(W\) whose edge restriction is divisible by \(p_{ij}\), because all summands vanish at its roots. The quotient vector has squared norm \(g^s\) on that edge. Thus the Gram matrix of the unit vectors \(W(u_i)\) satisfies the test with \(\eta_{ij}=\varepsilon_{ij}\sigma_{ij}^s\). All sign patterns arise in this way.

For sufficiency, take an extreme form and choose the square roots \(p_{ij}\) on its nonzero edges. At each edge with two positive vertex values the signs of the two endpoint values of \(p_{ij}\), together with \(\sigma_{ij}^s\), prescribe \(\eta_{ij}\). Extend these signs arbitrarily to the other edges. A feasible correlation matrix gives vertex vectors with the required normalized inner products. Scale them by \(\sqrt{f(u_i)}\), making a zero vector at every zero vertex. On each nonzero edge, choose a degree-\(s\) vector realizing the correlation from the interval lemma and multiply it by \(p_{ij}\). When a vertex value is zero, any normalized quotient works, as the endpoint inner product must be zero. Use the zero vector on a zero edge. After padding the vectors with zeros, an orthogonal map on each edge sends its endpoint vectors to the prescribed vertex vectors, because they have the same two-by-two Gram matrix. The components then glue to forms on \(X\), and their squares sum to \(g^s f\). The compact base now gives the inclusion for all nonnegative forms. Diagonal sign congruence on the correlation matrix switches all incident edge signs at a vertex. Hence one can fix the signs on a spanning forest and test only the remaining signs.

\textbf{The shortest-cycle formula.} The projective angles are metric distances. On a cycle, each edge length is at most the sum of the other lengths. If a cycle \(C\) has \(s\sum_{e\in C}\phi_e<\pi\), choose the edge signs there with product \(-1\). The interval test forces the spherical distance between the two vertex vectors, after the designated sign change, to be at most \(s\phi_e\). After following the cycle, accumulating signs, one joins a unit vector to its negative in length less than \(\pi\). This is impossible on a sphere, so the angular bound on every cycle is necessary.

Assume the cycle bounds and put \(\alpha_e=\min(s\phi_e,\pi/2)\). In every cycle \(\sum_e\alpha_e\ge\pi\). When at least two values are capped their contributions reach \(\pi\). When just one is capped, the uncapped remainder is at least the uncapped value of that edge and hence at least \(\pi/2\). The uncapped case is the assumed cycle bound. For any sign choice, prescribe \(B_{ij}=\eta_{ij}\cos\alpha_{ij}\) on the edges. These prescribed entries belong to the required signed intervals. For \(s=1\) this uses \(\phi_e\le\pi/2\). If \(\beta_e=\arccos B_{ij}\), then \(\alpha_e\le\beta_e\le\pi-\alpha_e\), so for every cycle \(C\) and odd subset \(F\subseteq C\), \[
\sum_{e\in F}(\pi-\beta_e)+\sum_{e\in C\setminus F}\beta_e\ \ge\ \pi.
\] These are Laurent's angle inequalities \cite{cmp11} for positive semidefinite completion on graphs with no \(K_4\) minor, and yield a correlation matrix. The completed matrix can be taken to have rank at most three \cite{cmp12}. The angle inequalities are vacuous for forests, which reach equality at level zero. For an arbitrary graph, at any \(s\ge2\) with \(s\phi_e\ge\pi/2\) for every edge, the identity matrix is feasible for every signing. This gives the asserted finite exact level in general.

A complete graph on \(n\ge3\) independent vertices with a common projective angle \(\theta\) has first exact level \(\lceil\arccos(1/(n-1))/\theta\rceil\). For the upper bound prescribe the entries \(\eta_{ij}\cos(\min(s\theta,\pi/2))\) and use diagonal dominance. At \(s=1\) the cap makes no change. For the lower bound choose all signs negative. Below the proposed level, all pairwise inner products would be at most \(-\cos(s\theta)<-1/(n-1)\), forcing the squared norm of the sum of the vertex vectors to be negative.

\textbf{Consequences for PPT extension.} The indicated nonedge quadrics and square-free cubics generate \(I(X)\), so \hyperlink{support-prop}{Proposition~3.1} applies for \(d\ge3\). It turns the correlation criterion into equality of the level-\(s\) cone and the separable cone. On failure of the criterion the proposition gives an entangled level-\(s\) marginal of rank \(\dim R_d=n+(d-1)|E(G)|\). The same construction gives full one-party rank because \(I(X)\) has no linear equations.

For any \(s\ge1\), choose an equiangular triangle with projective angle \(\theta\) satisfying \(\pi/(3(s+1))\le\theta<\pi/(3s)\). Its first exact level is \(s+1\). This gives a rank-\(3d\) ternary state passing level \(s\) and failing level \(s+1\). Adjoin \(N-3\) positive coherent rays on orthogonal complementary axes to obtain local rank \(N\) and operator rank \(3d+N-3\). Any extension would inherit the mixed quadrics between these blocks from the two-party marginal. A mixed monomial contains one of these quadrics, so the products in the apolar ideal force all mixed moments to vanish. Each block is positive, and contraction preserves its support. An extension at level \(s+1\) would therefore extend the original ternary block at that level, a contradiction. The lower bound of \hyperlink{res-1A}{Theorem~1A} now gives sharpness.

\Needspace{6\baselineskip}
\subsection{3.1 Finite exactness on plane cubics}\label{s-3-1}

\begin{paperresult}{Theorem 1D}{res-1D}
Let \(X\subset\mathbb P^2\) be a reduced real plane cubic such that each geometric component has Zariski-dense real points. Let \(g\) be the positive definite quadratic form used for contraction. For each \(d\ge3\), the PPT extension hierarchy on \(X\) equals the separable cone at some finite level exactly when every geometric singularity of \(X\setminus V(g)\) is a split real ordinary node. The smooth case satisfies this condition. When the singularity condition fails, there are entangled marginals of rank \(3d\) supported on that fixed cubic and admitted at every prescribed finite level, with the marginal allowed to depend on the level.

\end{paperresult}

Let \(X\) and \(g\) be as in \hyperlink{res-1D}{Theorem~1D}, and put \(U=X\setminus V(g)\). The affine chart \(U=D_+(g)\) has coordinate ring \((R[g^{-1}])_0\), and \(U(\mathbb R)=X(\mathbb R)\) is compact. Thus every homogeneous \(f\in P_{2d}(X)\) gives a nonnegative function \(f/g^d\) on \(U\). Conversely, a bound on the denominators of squares representing all these functions gives a single extension level. In detail, suppose the summands can be chosen from a fixed finite-dimensional subspace \(W\) of \(\mathbb R[U]\). Choose a common denominator \(g^k\) for a basis of \(W\) and increase \(k\) to make \(2k\ge d\). The representation of \(f/g^d\) then gives \[
g^{2k-d}f=\sum_i p_i^2,\qquad p_i\in R_{2k}.
\] The identity holds in \(R\) since \(g\) is a nonzero divisor. Indeed, \(g\) does not vanish identically on any geometric component, which has dense real points, and \(R\) is reduced. \hyperlink{support-prop}{Proposition~3.1} gives exactness at level \(2k-d\). If \(g^s f\) fails the square test for some nonnegative \(f\), the same proposition gives an entangled marginal with rank \(\dim R_d=3d\). Its proof applies to the cubic because \(I(X)\) has a single generator of degree three.

\textbf{Smooth curves and split nodes.} If \(U\) is smooth, apply Scheiderer's saturation and stability results \cite[Theorem 4.1 and Corollary 4.4]{cmp13} to \(U(\mathbb R)\) on each component. Saturation says that nonnegative functions are sums of squares. Stability means that for target functions in a fixed finite-dimensional subspace, one fixed finite-dimensional space of summands suffices. Apply it here to the space of \(f/g^d\). This also proves finite exactness on any smooth projective curve with Zariski-dense real points on every component, on the relative Hankel cone. The same statement applies to ambient CS extensions when the defining ideal is generated through degree \(d\).

Suppose instead that all geometric singularities of \(U\) are split real ordinary nodes. Let \(\widetilde U\) be the normalization. Its real locus is compact, since it is the full real locus of the projective normalization of \(X\). The ring \(\mathbb R[U]\) consists of regular functions on \(\widetilde U\) agreeing at the two preimages of each node: at a split node the completed local ring consists of pairs of series with the same constant term \cite[Lemma 1.6]{cmp14}, and the completion detects the cokernel of the finite normalization map. Applying stability on \(\widetilde U\) first gives a vector \(P\) of summands representing the pullback of \(f/g^d\), from a fixed finite-dimensional space. We show that the summands can be made to agree at all the pairs of preimages while keeping a uniform bound on their spaces.

For a pair \(a,b\) of preimages, choose a regular function \(h\) on \(\widetilde U\) with \(h(b)=1\) and \(h=0\) at every other preimage, by the Chinese remainder theorem. Put \(u=(h-1)^2\) and choose \(B>\max(2,\|u\|_\infty)\) on the compact real locus. Choose an integer \(k\ge0\) with \((B-1)(1-1/B)^k\le1\). The function \(A=(1-u)(1-u/B)^k\) has \(|A|\le1\). This follows on \(0\le u\le1\) from both factors lying in \([0,1]\), and on \(1\le u<B\) from the choice of \(k\). Therefore \(F=1-2A^2\) has \(|F|\le1\), with \(F(b)=-1\) and \(F=1\) at every other preimage. By saturation on the smooth normalization, fix a representation \(1-F^2=\sum_j H_j^2\). Every \(H_j\) vanishes at every preimage.

If \(P(a)=P(b)\), the pair already agrees. Otherwise put \(e=(P(a)-P(b))/\|P(a)-P(b)\|\), and decompose \(P\) into the scalar component \(p=e^{\mathsf T}P\) and \(P_\perp=P-ep\). Equality of the norms of \(P(a)\) and \(P(b)\) gives \(p(a)=-p(b)\), while \(P_\perp(a)=P_\perp(b)\). Replace \(P\) by the vector with first block \(P_\perp+eFp\) and extra entries \(H_jp\). Its squared norm is \(\|P_\perp\|^2+(F^2+\sum_j H_j^2)p^2=\|P\|^2\). Its values agree at the chosen pair. At every other preimage it equals the old vector with zeros appended. For a fixed input space \(W_i\), all output entries lie in \(W_{i+1}=W_i+FW_i+\sum_j H_jW_i\), independently of \(P\). Repeat for every node and intersect the final fixed finite-dimensional space with \(\mathbb R[U]\). This supplies the uniform summand space and proves finite exactness under the stated singularity condition.

\textbf{Obstructions on the other plane cubics.} In each case below a real quadratic \(q\) is nonnegative on \(X(\mathbb R)\) and \(g^{d-1}q\) supplies a witness at every finite level. For an irreducible singular cubic the singular point is unique, hence real. Indeed a line through two singular points would exceed its allowed intersection number with the cubic. Take it to be \([0:0:1]\) and write the equation as \(zQ_2(x,y)+Q_3(x,y)=0\), with \(Q_2\ne0\). The indefinite rank-two case is a split node. If \(Q_2\) has rank one, arrange \(Q_2=y^2\). Irreducibility forces a nonzero \(x^3\) term in \(Q_3\). A shear of \(x\) kills the \(x^2y\) term, and a change of \(z\) by a linear form kills the remaining terms divisible by \(y^2\). If \(Q_2\) is definite, arrange \(Q_2=x^2+y^2\). Modulo multiples of \(Q_2\), a cubic is a linear combination of the real and imaginary parts of \((x+iy)^3\). A rotation and a change of \(z\) by a linear form then give the ordinary solitary node. This gives the following two forms and witnesses.

\begin{itemize}
\tightlist
\item
  For the cusp \(y^2z=x^3\), take \(q=xz\). On \(z=1\) one has \(x=t^2\), \(y=t^3\) near the cusp, so \(q\) has order two in \(t\). A square summand in a proposed representation must vanish at the cusp. Each such summand has order at least two before squaring, since functions in the local ring have no linear term in \(t\). A sum of their squares has order at least four.
\item
  For the solitary node \(y^2z=x^2(x-z)\), again take \(q=xz\). On \(z=1\) the real points have either \(x=0\) or \(x\ge1\), so the witness is nonnegative. At the node \(x\) has nonzero class in \(\mathfrak m/\mathfrak m^2\), since the equation has no linear part. A sum of squares vanishing at the real point belongs to \(\mathfrak m^2\).
\end{itemize}

For a line and an irreducible conic with real points, two transverse real intersections are split nodes. For tangency or conjugate intersections use the following projective normal forms. The tangent case is obtained by taking the double point of intersection to \([0:0:1]\) and its tangent to \(y=0\). In the conjugate case, with the line at \(z=0\), the restriction of the conic to that line is definite. Completing the square and rescaling gives \(x^2+y^2=z^2\).

\begin{itemize}
\tightlist
\item
  For the line \(y=0\) tangent to \(yz=x^2\), take \(q=yz\). It is zero on the line and nonnegative on the conic. At the tangency point \(q\) has nonzero class in \(\mathfrak m/\mathfrak m^2\) in the chart \(z=1\), again impossible for a sum of squares vanishing there.
\item
  For the line \(z=0\) and the conic \(x^2+y^2=z^2\), take \(q=z(z+\epsilon x)\) with \(0<\epsilon<1\). This is zero on the line and nonnegative on the conic. Any square summand must vanish identically on the line, so its restriction to the conic vanishes at least to first order at their two conjugate intersections. Every square then vanishes at least to second order. The restriction of \(q\) has order one there. This obstruction applies when the intersections lie in \(U\). When \(g\) vanishes at them, \(U\) is the disjoint union of smooth components and finite exactness follows from the smooth case.
\item
  For three concurrent real lines, arrange that the lines have equations \(x=0\), \(y=0\) and \(x=y\), and take \(q=xy\). It is nonnegative on the union. Each proposed square summand must vanish identically on the first two lines and therefore be divisible by \(xy\). On the third line its square has order at least four at the common point, whereas \(q\) has order two. Three nonconcurrent lines have only split nodes.
\end{itemize}

In all these obstructions the asserted order or class is preserved by multiplication by any power of \(g\), since \(g\) is nonzero at the point used. The witness remains nonnegative on the full real curve, including points at infinity. A reduced plane cubic with every geometric component real and totally real is either irreducible, a line with an irreducible conic, or three real lines. This exhausts the alternatives and proves \hyperlink{res-1D}{Theorem~1D}.

For example, on \(z(x^2+y^2-z^2)=0\) the complex intersections are \([1:\pm i:0]\). Contraction by \(g=x^2+y^2+z^2\) gives finite exactness, while \(g=2x^2+y^2+z^2\) gives nonexactness at every finite level.

For a finite-exactness case with real algebraic coefficients, its first exact level is computable in principle by real quantifier elimination. For each \(s\), ask whether every degree-\(2d\) form nonnegative on the real unit representatives of \(X\) for \(g\) has a positive semidefinite Gram representation of \(g^s f\) in \(R_{2(d+s)}\). This is a first-order sentence using the equations of \(X\), the entries of the Gram matrix and its principal minors. Testing levels in increasing order terminates by the preceding existence proof.

\Needspace{6\baselineskip}
\subsection{3.2 Exact levels on split-nodal cubics}\label{s-3-2}

\begin{paperresult}{Theorem 1E}{res-1E}
Let \(X\) be a geometrically irreducible real plane cubic with a split real node and let \(\nu:\mathbb P^1\to X\) be its normalization by binary cubics. Choose real lifts \(a,b\) of the two preimages of the node whose images under the cubics are the same \(g\)-unit vector. Write \(G=g\circ\nu\). For every \(d\ge3\), the level-\(s\) PPT extension cone on \(X\) equals the separable cone if and only if a real binary form \(A\) of degree \(3s\) satisfies \[
A(a)=1,\qquad A(b)=-1,\qquad A^2\le G^s\quad\text{on }\mathbb R^2.
\] This is one semidefinite feasibility test of matrix size \(3s+2\). The first exact level depends on \((X,g)\) and is independent of \(d\). \hyperref[s-3-2]{Section~3.2} gives an explicit upper bound and examples with first exact levels one and two. As the projective embedding varies within the class of a fixed such cubic, the first exact level is unbounded.

\end{paperresult}

Let \(X\) be geometrically irreducible with a split ordinary node. Projection from the node is birational to \(\mathbb P^1\), with the two node preimages corresponding to the two real tangent lines. The normalization is given by three binary cubics \(\nu_i\). Indeed, with node \([0:0:1]\) and equation \(zQ_2(x,y)+Q_3(x,y)=0\), the third point on the line of direction \([u:v]\) has coordinates \([uQ_2(u,v):vQ_2(u,v):-Q_3(u,v)]\). These cubics have no common root, since a common root of \(Q_2,Q_3\) would give a line component. Write \(\nu\) also for the vector of these cubic forms on \(\mathbb R^2\). Its odd degree allows real rescaling of the lifts \(a,b\) so that \(\nu(a)=\nu(b)\) is a \(g\)-unit vector. Thus \(G=g\circ\nu\) is a strictly positive binary sextic and \(G(a)=G(b)=1\).

For \(j\ge1\) the pullback identifies \(R_j\) with the binary forms \(P\) of degree \(3j\) satisfying \(P(a)=P(b)\). Indeed, pullback is injective and its image has dimension \(3j\), the plane cubic Hilbert function. The equality at \(a,b\) is a single independent constraint on binary forms of degree \(3j\), a space of dimension \(3j+1\). Nonnegativity on \(X(\mathbb R)\) is equivalent to nonnegativity after pullback, since the split node and all smooth real points have real preimages.

\textbf{Reduction to binary forms.} The cone of nonnegative pullbacks in degree \(6d\) has a compact base by summing finitely many real evaluations spanning the dual of \(R_{2d}\). Each extreme ray is a square \(p^2\) of a real binary form of degree \(3d\) with only real roots. To see this, factor a nonzero nonnegative form as \(h^2 q\), gathering all real zeros in \(h\). If the strictly positive factor \(q\) has positive degree, its degree is at least two. Choose a nonzero form of that degree vanishing at \(a,b\) and use it to perturb \(q\) by sufficiently small amounts in both signs. The perturbation keeps the equality at the preimages and gives a nontrivial decomposition. Thus \(q\) is constant on an extreme ray.

Suppose the certificate \(A\) in \hyperlink{res-1E}{Theorem~1E} exists. Then the nonnegative binary form \(G^s-A^2\) is a sum of squares of degree-\(3s\) binary forms, by factorization over \(\mathbb C\). These summands vanish at \(a,b\). With \(A\) they form a vector \(Q\) whose squared norm is \(G^s\) and whose values at \(a,b\) are opposite unit vectors. For an extreme ray \(p^2\) with \(p(a)=-p(b)\), the components of \(pQ\) have equal values at \(a,b\) and descend to \(R_{d+s}\). If \(p(a)=p(b)\), use instead the degree-\(3s\) vector obtained by taking the \(s\)-fold tensor power of a linear change of the \(\nu_i\) representing \(G\) as a sum of squares. It has identical values at \(a,b\). This also covers zero endpoint values. The resulting representations for the extreme rays, and then their sums, yield \(g^sP_{2d}(X)\subseteq\Sigma_{2(d+s)}(X)\).

For necessity, take \(p\) of degree \(3d\) with distinct real roots away from \(a,b\) and with \(p(a)=1\), \(p(b)=-1\). In coordinates dual to \(a,b\) one can take a product of linear factors with distinct nonzero real roots and prescribe their product. The square descends to a nonnegative form on \(X\). In a proposed SOS representation of \(G^s p^2\) by pullbacks from \(R_{d+s}\), every summand is divisible by \(p\), since it vanishes on its distinct real roots. The quotient vector \(Q\) has squared norm \(G^s\) and opposite unit values at \(a,b\). Its scalar projection onto \(Q(a)\) satisfies the certificate. Thus failure of the binary condition gives a nonnegative form without an SOS representation at that level. \hyperlink{support-prop}{Proposition~3.1} identifies this SOS inclusion with exactness of the PPT cone on \(X\). A failure gives entangled rank-\(3d\) marginals.

Let \(m_s\) be the vector of binary monomials of degree \(3s\) and write \(A=c^{\mathsf T}m_s\). An equivalent formulation of the inequality \(A^2\le G^s\) is \[
m_s^{\mathsf T}B m_s=G^s,\qquad
\begin{pmatrix}1&c^{\mathsf T}\\ c&B\end{pmatrix}\succeq0 .
\] One direction follows by taking a Gram matrix for the nonnegative binary residual. The other follows by the Schur complement. The two endpoint constraints are linear in \(c\). The matrix has size \(3s+2\) and the problem depends on \((X,g)\) and \(s\) alone. In particular, level zero is never exact on this curve.

\textbf{An explicit upper bound.} Choose real linear forms \(\lambda,\mu\) dual to \(a,b\), so \(L=\lambda^6-\mu^6\) takes values \(1,-1\) at the preimages. If \(\gamma\) is the least value of \(G\) on the Euclidean unit circle of the parameter plane, then \[
\left|\frac{L}{G}\right|\le B_0
=\max\!\left(1,\frac{\|\lambda\|^6+\|\mu\|^6}{\gamma}\right).
\] Any positive lower bound for \(\gamma\) can be used. Take an odd integer \(m\ge\pi B_0/2\) and set \(\tau=\sin(\pi/(2m))\). Then \(|\tau L/G|\le1\), and \(T_m(\tau)=(-1)^{(m-1)/2}\) for the Chebyshev polynomial \(T_m\). The homogeneous binary form \[
A=(-1)^{(m-1)/2}G^m T_m(\tau L/G)
\] is of degree \(6m\). It takes values \(1,-1\) at the preimages and satisfies \(A^2\le G^{2m}\). The first exact level is therefore at most \(2m\). For real algebraic coefficients all these ingredients can be computed or bounded by univariate algebraic inequalities.

\textbf{First exact levels one and two.} Use the Euclidean \(g\) and the family \[
\nu_{\alpha,\beta}(u,v)=
\big(\alpha(u^2-v^2)v,\ \beta(u^2-v^2)u,\ v^3\big),
\qquad \alpha,\beta>0 .
\] The node preimages are \((1,1)\) and \((-1,1)\). On the chart \(v=1\), put \(t=u/v\). Here \[
G(t,1)=1+(t^2-1)^2(\alpha^2+\beta^2t^2).
\] At level one, any feasible certificate can be replaced by its odd part in \(t\), since \(G\) is even and the endpoint conditions are opposite. Its degree is at most three. At \(t=1\) the nonnegative difference \(G-A^2\) has a zero, so \(A'(1)=0\), using \(G'(1)=0\) and \(A(1)=1\). These conditions uniquely give \(P(t)=(3t-t^3)/2\). Directly, \[
G(t,1)-P(t)^2=(t^2-1)^2\big(1+\alpha^2+(\beta^2-1/4)t^2\big).
\] Thus level one is exact precisely when \(\beta\ge1/2\). For general \(s\ge1\), homogenize this same polynomial to degree \(3s\) using the factor \(v^{3s-3}\). The inequality \((1+w)^s\ge 1+sw\) for \(w\ge0\) gives \[
G(t,1)^s-P(t)^2\ \ge\
(t^2-1)^2\big(1+s\alpha^2+(s\beta^2-1/4)t^2\big).
\] Consequently \(s_\star\le\lceil1/(4\beta^2)\rceil\) in this family. With \(\alpha=1\), \(\beta=1/2\) the first exact level is one. With \(\alpha=1\), \(\beta=2/5\) it is two. These are rational examples with a negative test at the preceding level for every \(d\ge3\).

\textbf{Prescribing the first exact level on a fixed nodal cubic.} Keep \(X\) and its normalization fixed and choose parameter coordinates \((t,u)\) with the node lifts at \((1,0),(0,1)\). The normalization components span the three-dimensional space of cubics whose endpoint values agree, so they are related by a fixed invertible linear change to the basis \[
e=(t^3+u^3,\ t^2u,\ tu^2)^{\mathsf T}.
\] For \(0\le c<1\) put \(q_c=t^2+2ctu+u^2\) and \(\lambda=t+u\). The identity \[
q_c^3-\left(\frac{3\lambda q_c-\lambda^3}{2}\right)^2
=\frac{(q_c-\lambda^2)^2(4q_c-\lambda^2)}4
\] gives a positive definite Gram matrix for \(q_c^3\) on this basis. The cubic being squared on the left is \(e_1+3c(e_2+e_3)\). The right side is the positive definite quadratic \(4q_c-\lambda^2\) on the two linear coordinates, multiplied by \((1-c)^2t^2u^2\). Its diagonal coefficients are \(3\) and its off-diagonal coefficient as a quadratic matrix is \(4c-1\in[-1,3)\). The resulting two squares span the space of cubics divisible by \(tu\). Together with the first cubic this gives a rank-three Gram matrix. It is the pullback of a positive definite \(g_c\) on the fixed plane, and \(g_c\circ\nu=q_c^3\).

By the endpoint interval calculation of \hyperref[s-3]{Section~3}, the endpoint inner products of vectors with squared norm \(q_c^{3s}\) range over \(J_{3s}(\arccos c)\). The split-node condition is that this interval contain \(-1\), because the quotient vector must have opposite unit values at the two preimages. Thus level \(s\) is exact if and only if \(3s\arccos c\ge\pi\). Level zero is inexact, so this proves the formula in \hyperlink{res-1H}{Corollary~1H} for geometrically irreducible nodal cubics.

\Needspace{6\baselineskip}
\subsection{3.3 Exact levels on smooth plane cubics}\label{s-3-3}

\begin{paperresult}{Theorem 1F}{res-1F}
Let \(X\) be a smooth real plane cubic and \(g\) the positive definite quadratic form used for contraction. In real projective coordinates put \(X\) in Weierstrass form \[
y^2z=x^3+A x^2z+B xz^2,
\] with \(0\) the smallest real root of \(x(x^2+Ax+B)\). In particular \(B>0\). For every \(d\ge3\) and \(s\ge1\), the level-\(s\) PPT hierarchy on \(X\) is exact if and only if \(g^s xz\) is a sum of squares on \(X\). This is one Gram feasibility test of size \(3s\). For \[
g=a x^2+b y^2+c z^2+2p xy+2q xz+2r yz,
\] level one is exact if and only if the following matrix is positive semidefinite: \[
Q_1(g)=
\begin{pmatrix}
 b & (a+Ab-c/B)/2 & p\\
 (a+Ab-c/B)/2 & Bb+2q-Ac/B & r\\
 p & r & c/B
\end{pmatrix}.
\] At every exact level \(g^s xz\) is a sum of at most three squares on the cubic. For each fixed \(X\), the first exact level is independent of \(d\), and every positive integer occurs as \(g\) varies. If level \(s\) is exact for \(g\), then the level-\((s+1)\) Gram test is strictly feasible and remains exact in a neighborhood of \(g\) among positive definite quadratics.

\end{paperresult}

We prove \hyperlink{res-1F}{Theorem~1F} in the homogeneous coordinate ring. Put \(\mathcal L=\mathcal O_X(1)\). The degree-\(j\) ring is \(H^0(X,\mathcal L^j)\) for \(j\ge1\), by the exact sequence for the cubic in \(\mathbb P^2\) and the vanishing of \(H^1(\mathbb P^2,\mathcal O(j-3))\). The nonnegative real sections of \(\mathcal L^{2d}\) are defined by the sign in the square of each real fiber of \(\mathcal L^d\).

\textbf{Reduction to real zero divisors.} Let \(f\) be a nonzero nonnegative section of \(\mathcal L^{2d}\), with real zero divisor \(2D\). If \(f\) has nonreal zeros, then \(\deg D<3d\). Riemann--Roch on the genus-one curve gives a nonzero real \(u\in H^0(X,\mathcal L^d(-D))\). On the real curve the function \(u^2/f\) extends across the zeros by the prescribed orders of vanishing and attains a positive maximum \(c\). The section \(f-u^2/c\) is nonnegative and vanishes at all old real zeros to at least their original orders. At a real maximum point it has an additional zero, or a higher order at an existing zero. Every zero of a nonzero nonnegative section at a smooth real point has even order. Iterating gives \(f\) as a sum of squares of degree-\(d\) sections plus a nonnegative remainder whose divisor is entirely real, or zero. The number of iterations is bounded by \(3d\), since at each step the total real zero multiplicity increases by at least two. This is the relative homogeneous version of the reduction by real zeros in \cite[Lemma 2.7]{cmp15}.

\textbf{Reference forms and the exact test.} A nonzero remainder as above has divisor \(2D\), with \(\mathcal O_X(D)=\mathcal L^d\otimes T\) for a real two-torsion line bundle \(T\). There are at most four such real classes, as the complex elliptic curve has four two-torsion classes. To construct a real effective divisor \(D_T\) in the class \(\mathcal L\otimes T\), choose two real points, with repetition allowed. Subtracting them from the class leaves a real line bundle of degree one. By Riemann--Roch it has a one-dimensional real space of sections, and its zero divisor is a real point. Taking all three points gives the required \(D_T\). A square of a real section of \(\mathcal L\otimes T\), using a real trivialization of \(T^2\), gives \(f_T\) of divisor \(2D_T\).

For any remainder \(f\) of class \(T\), there is a real meromorphic section \(h\) of \(\mathcal L^{d-1}\) with divisor \(D-D_T\), since that divisor represents \(\mathcal L^{d-1}\). The quotient of \(f\) by \(h^2 f_T\) is a nonzero real constant. In particular, a class can occur among the nonnegative remainders only if an overall sign makes \(f_T\) nonnegative on all real components. After choosing the sign, absorb a positive scalar into \(h\) to obtain \(f=h^2 f_T\).

If \(g^s f_T=\sum_i b_i^2\), each section \(b_i\) vanishes along \(D_T\). Indeed every point of \(D_T\) is real, and the smallest order among the summands controls the order of the sum of their squares. Also \(g\) has no real zeros. Hence each \(h b_i\) is a regular section of \(\mathcal L^{d+s}\), and they give an SOS for \(g^s f\). The squares subtracted in the reduction are also SOS after multiplication by \(g^s\), since \(g\) is itself a sum of squares of linear forms. This proves sufficiency for every \(d\). For necessity, choose a real linear section \(\ell\) with all its zeros real, as in the construction of \(D_T\) for the trivial class. If \(g^s f_T\ell^{2(d-1)}\) is SOS, each summand is divisible on \(X\) by \(\ell^{d-1}\), again by the vanishing orders at real points. Dividing yields an SOS for \(g^s f_T\). Thus the level-\(s\) inclusion in degree \(2d\) is equivalent to these degree-two reference tests. By \hyperlink{support-prop}{Proposition~3.1} this is equivalent to PPT exactness for \(d\ge3\).

\textbf{The Gram matrix on the vanishing space.} The same order-of-vanishing argument restricts every sum-of-squares representation of \(g^s f_T\) to the real sections of \[
V_{T,s}=H^0(X,\mathcal L^{s+1}(-D_T))\simeq H^0(X,\mathcal L^s\otimes T^{-1}).
\] For \(s\ge1\) this line bundle has degree \(3s\), and its dual has no sections. Riemann--Roch gives \(\dim_{\mathbb R}V_{T,s}=3s\). If \(b_1,\ldots,b_{3s}\) is a real basis, the test is to find \(Q\succeq0\) satisfying \(\sum_{i,j}Q_{ij}b_i b_j=g^s f_T\) modulo the cubic equation. A basis is obtained in \(R_{s+1}\) by imposing vanishing of derivatives in a real local parameter through order \(m-1\) at each point of multiplicity \(m\) in \(D_T\). These are three independent linear conditions, including when points repeat. At \(s=0\) the vanishing space is zero for every nontrivial \(T\), so a retained nontrivial class obstructs exactness at level zero. The trivial class is a square and needs no test. The following choice avoids having to compute divisors or signs separately.

\textbf{The single reference class on a real Weierstrass cubic.} Choose a real flex \(O\). The tangent at \(O\) cuts \(3O\), and the real plane cubic admits real Weierstrass coordinates \cite{cmp16} with \(O=[0:1:0]\). Let \(P_e=[e:0:1]\) for each real root \(e\) of the affine monic cubic. These are the nonidentity real two-torsion points since inversion changes the sign of \(y\). The line \(x-ez=0\) cuts \(2P_e+O\), and \(z=0\) cuts \(3O\). Therefore \((x-ez)z\) has divisor \(2(P_e+2O)\), representing the class of \(P_e-O\) in the reference test. On the affine real locus its sign is that of \(x/z-e\). The smallest real root gives a globally nonnegative form. If there are three real roots, each other root gives a sign change between a point with \(x/z\) strictly between the two smallest roots and a point with \(x/z\) above the largest root. Thus precisely one nontrivial class has nonnegative orientation. Shift this smallest root to zero. The derivative of the monic cubic there is \(B>0\), and the single reference form is \(xz\), with half divisor \(P_0+2O\). The nontrivial class also rules out level zero.

\textbf{The level-one matrix.} In degree two the vanishing conditions at \(P_0\) and \(2O\) give the basis \(u=(x^2,xz,yz)^{\mathsf T}\) of \(V_{T,1}\). The six products of basis elements are independent modulo the cubic. Indeed a linear multiple of the cubic containing \(x\) or \(y\) in its multiplier has an \(xy^2z\) or \(y^3z\) term outside their span. A multiple by \(z\) has an \(xz^3\) term with nonzero coefficient \(B\), also outside their span. The products before taking the quotient are distinct monomials. Since \(y^2z^2=x^3z+A x^2z^2+B xz^3\), expansion yields \(g xz=u^{\mathsf T}Q_1(g)u\) on \(X\), with \(Q_1(g)\) as in \hyperlink{res-1F}{Theorem~1F}. Independence gives uniqueness, so a principal-minor test decides exactness without solving an SDP.

\Needspace{5\baselineskip}
\hypertarget{smooth-univariate}{}\textbf{A univariate form of the smooth test.} Write \(p_2(x)=x^2+Ax+B\) and work on \(z=1\), where \(y^2=x p_2(x)\). Reduce \(g(x,y,1)^s\) to its unique normal form \(H_s(x)+yJ_s(x)\). Define \[
r_s=\left\lfloor\frac{3s-1}{2}\right\rfloor,\qquad
t_s=\left\lfloor\frac{3s-2}{2}\right\rfloor.
\] For \(s\ge1\), level \(s\) is exact if and only if there are real univariate polynomials \(U,V\) with \(\deg U\le2r_s\), \(\deg V\le2t_s\), satisfying \[
H_s=xU+p_2V,\qquad
M(x)=\begin{pmatrix}U(x)&J_s(x)/2\\ J_s(x)/2&V(x)\end{pmatrix}\succeq0
\quad\text{for every }x\in\mathbb R.
\] Indeed, dividing sections of \(\mathcal L^{s+1}\) by \(z^{s+1}\) identifies them with affine regular functions having a pole of order at most \(3(s+1)\) at \(O\). The extra conditions from the half divisor \(P_0+2O\) give vanishing at \(P_0\) and pole order at most \(3s+1\). In the normal form \(a(x)+y b(x)\), the pole orders of the two terms have different parity, since \(x\) and \(y\) have pole orders two and three at \(O\). Thus the admissible functions are exactly \(x u(x)+y v(x)\) with \(\deg u\le r_s\) and \(\deg v\le t_s\). Expanding their squares and dividing by \(x\) gives the test, with \(U=\sum u_i^2\), \(V=\sum v_i^2\) and \(J_s=2\sum u_iv_i\).

Conversely the real univariate matrix factorization theorem gives \(M=Q^{\mathsf T}Q\) with a real polynomial matrix \(Q\) of size at most \(3\times2\) \cite[Corollaries 2.7 and 2.8 in the preprint]{cmp19}. The diagonal entries, which are sums of squares, bound the degrees in the two columns by \(r_s\) and \(t_s\) respectively. Hence the rows of \(Q\) give at most three square summands in \(V_{T,s}\). This bound is sharp uniformly in \((X,g,s)\): on \(y^2z=x^3+xz^2\) with \(g=x^2+y^2+z^2\), the unique level-one Gram matrix is the identity of order three. If \(g\) is invariant under \(y\mapsto-y\), then \(J_s=0\). The condition reduces to scalar nonnegativity of \(U\) and \(V\) on \(\mathbb R\), with two Gram blocks of orders \(\lceil3s/2\rceil\) and \(\lfloor3s/2\rfloor\).

\textbf{Squares in the two-torsion line bundle.} Let \(T\) be the retained nontrivial class and choose a section \(\tau\) of \(\mathcal L\otimes T\) with divisor \(P_0+2O\). Trivialize \(T^2\) so that \(\tau^2=xz\). This trivialization makes squares nonnegative on every real fiber, since \(xz\) is nonnegative and positive off a finite set on the real curve. Write \(U_s=H^0(X,\mathcal L^s\otimes T)\) and \(\Sigma_s^T\) for the cone of sums of squares of its real sections in \(H^0(X,\mathcal L^{2s})\), with products using that trivialization. The vanishing argument above and division by \(\tau\) identify the test \(g^s xz\in\Sigma_{2s+2}(X)\) with \(g^s\in\Sigma_s^T\). Each \(\Sigma_s^T\) is closed. Finitely many real evaluations can be chosen to span \(U_s^*\) after choosing local frames, by real density. A sum of these evaluations on the squared sections bounds the trace of every positive semidefinite Gram matrix representing a convergent sequence in the cone.

\textbf{Strict feasibility after one level.} More generally, let \(w=\sum_i a_i^2\in\Sigma_s^T\) be strictly positive on \(X(\mathbb R)\), with \(s\ge1\), and let \(h\) be any positive definite quadratic. We show that \(hw\) belongs to the interior of \(\Sigma_{s+1}^T\) in \(H^0(X,\mathcal L^{2s+2})\).

If \(hw\) is on the boundary, there is a nonzero real linear functional \(\Lambda\) nonnegative on squares from \(U_{s+1}\) with \(\Lambda(hw)=0\). Write \(h=\sum_{j=1}^3\ell_j^2\) for a basis of real linear forms. The pairing \((v_1,v_2)\mapsto\Lambda(v_1v_2)\) on \(U_{s+1}\) is positive semidefinite. The sum of its diagonal values on the \(a_i\ell_j\) is zero, so every \(a_i\ell_j\) lies in its kernel. Thus \(\Lambda\) kills all products of \(a_i\) with \(H^0(X,\mathcal L^{s+2}\otimes T)\). For this last assertion, choose two linear forms without a common zero on \(X\). They give a surjection \(\mathcal O_X^2\to\mathcal L\) with kernel \(\mathcal L^{-1}\). After tensoring with \(\mathcal L^{s+1}\otimes T\), the kernel has positive degree and vanishing \(H^1\), so products of linear forms with \(U_{s+1}\) span \(H^0(X,\mathcal L^{s+2}\otimes T)\).

Let \(B\) be the common zero divisor of the \(a_i\). It has no real support because \(w\) is positive, and \(\deg B\le3s\). With \(M=\mathcal L^s\otimes T(-B)\), the sections obtained by removing \(B\) span a base-point-free system in \(M\). If the system is one-dimensional, its generator has no zeros, so multiplication by that generator is an isomorphism onto the corresponding product space. Otherwise choose two real combinations without a common zero, by fixing one nonzero section and avoiding the finitely many conditions at its zeros. Their evaluation map \(\mathcal O_X^2\to M\) has kernel \(M^{-1}\). Tensor with \(\mathcal L^{s+2}\otimes T\). The kernel is \(\mathcal L^2(B)\), of positive degree, so the resulting map on global sections is surjective. Restoring the common factor \(B\), in either case the product space killed by \(\Lambda\) contains all of \(H^0(X,\mathcal L^{2s+2}(-B))\).

If \(B\) is empty this already contradicts \(\Lambda\ne0\). Otherwise \(\Lambda\) factors through restriction to \(B\). The restriction of \(U_{s+1}\) to \(B\) is surjective because \(\deg(\mathcal L^{s+1}\otimes T(-B))\ge3\), and the restriction of the target is surjective for the same reason. The complexified coordinate algebra of \(B\) is the product of its local Artinian algebras, exchanged in conjugate pairs because \(B\) has no real support. Assign the constants \(i\) and \(-i\) to each such pair. The resulting element is fixed by conjugation, hence descends to a real \(\iota\in H^0(B,\mathcal O_B)\) with \(\iota^2=-1\), even for a nonreduced divisor. Thus for any restricted section \(v\), both \(\Lambda(v^2)\) and \(\Lambda((\iota v)^2)=-\Lambda(v^2)\) are nonnegative. All squared sections on \(B\) are killed. They span the restricted target by polarization and a trivialization of the line bundle on the finite scheme. This contradicts \(\Lambda\ne0\) and proves interior membership. In particular there is a positive definite Gram matrix on \(U_{s+1}\): subtract a sufficiently small positive multiple of the sum of squares of any basis, represent the remainder in the cone, and add that multiple of the identity matrix back.

\textbf{Every first exact level on one smooth cubic.} A diagonal \(h\) with \(b=1\) and \((a,c)=(1-A,B)\) gives a positive definite \(Q_1(h)\) when \(A\le0\). If \(A>0\), the other roots cannot both be real, since their sum would be negative and their product positive, putting them below zero. Hence \(A^2<4B\). Choose \(a=t\), \(b=1\), \(c=Bt\) for \(t=(B-A^2/4)/(2A)>0\). The upper left \(2\times2\) block of \(Q_1(h)\) then has determinant \((B-A^2/4)/2\) and the remaining diagonal entry is \(t\), with zero off-diagonal entries in that row.

Now set \(g_\epsilon=z^2+\epsilon h\) for \(\epsilon>0\). For fixed \(s\), exactness is monotone in \(\epsilon\). If \(g_\epsilon^s\in\Sigma_s^T\), expand \((g_\epsilon+\delta h)^s\) for \(\delta>0\). Each added term contains a factor \(h\) that is a sum of squares from \(U_1\), and the remaining factors are ordinary sums of squares. Thus the full expansion lies in \(\Sigma_s^T\). For large enough \(\epsilon\), positive definiteness of \(Q_1(h)\) gives exactness at level one, since \(g_\epsilon/\epsilon\to h\). At the other end, \(z^{2s}\) cannot lie in \(\Sigma_s^T\). If it did, each summand would vanish to order at least \(3s\) at the real flex \(O\). Dividing by \(z^s\), which has divisor \(3sO\), would give a nonzero section of \(T\), impossible. Closedness therefore makes level \(s\) inexact for all sufficiently small positive \(\epsilon\).

For each \(s\), the exact values of \(\epsilon\) form a closed ray \([\epsilon_s,\infty)\) with \(0<\epsilon_s<\infty\). The strict feasibility result at \(g_{\epsilon_s}\) with multiplier \(g_{\epsilon_s}\) puts \(g_{\epsilon_s}^{s+1}\) in the interior of \(\Sigma_{s+1}^T\). By continuity, level \(s+1\) is already exact at some smaller \(\epsilon\), giving \(\epsilon_{s+1}<\epsilon_s\). Every fixed \(\epsilon>0\) has a finite exact level by \hyperlink{res-1D}{Theorem~1D}, so \(\epsilon_s\to0\). Level one first becomes exact for \(\epsilon\in[\epsilon_1,\infty)\), while for each \(s\ge2\) the interval \([\epsilon_s,\epsilon_{s-1})\) has first exact level \(s\). This proves the smooth case of \hyperlink{res-1H}{Corollary~1H}. Each such interval contains rational parameters. For real algebraic Weierstrass data, the choices above make \(h\) real algebraic, and the Gram feasibility conditions define a semialgebraic set in \(\epsilon\) over the real algebraic numbers. The endpoint \(\epsilon_s\) is then real algebraic and can be found by real quantifier elimination.

\Needspace{7\baselineskip}
\textbf{The dual pairing of a pencil without base points.} Let \(X\) be a smooth or geometrically integral split nodal plane cubic, let \(M\) be a real line bundle of degree \(m\ge3\), and let \(a,b\in H^0(X,M)\) have no common complex zero. There is a one-dimensional space of functionals on \(H^0(X,M^2)\) annihilating \(aH^0(X,M)+bH^0(X,M)\). For a nonzero element \(\Lambda\) of this line, the pairing \((v,v')\mapsto\Lambda(vv')\) has kernel exactly \(\langle a,b\rangle\).

By adjunction the dualizing sheaf is trivial in either case. Duality and Riemann--Roch \cite{cmp21} give \(h^0(N)=\deg N\) and \(h^1(N)=0\) for every positive-degree line bundle \(N\). The pencil evaluation sequence tensored with \(M\) has kernel \(\mathcal O_X\). The product map on global sections therefore has one-dimensional kernel spanned by \((b,-a)\), and its image is a hyperplane. Choose a general real member \(a\) of the pencil with simple zeros \(P_1,\dots,P_m\) in the smooth locus. The pencil maps to the projective line and the induced nonconstant map on the normalization has only finitely many branch values. Avoid these and the images of the singular points. Trivialize at the zeros by \(b\). Evaluation of \(H^0(X,M)\) has kernel \(\langle a\rangle\) and image a hyperplane \(W\subset\mathbb C^m\) containing \(\mathbf 1\). Its defining relation \(\sum_j c_jv_j=0\) has every \(c_j\ne0\), because evaluation at all but any one of these points is onto by vanishing for the degree-one twist. Evaluation of \(H^0(X,M^2)\) at all \(m\) points is onto with kernel \(aH^0(X,M)\), by the degree-\(m\) vanishing. In the trivialization by \(b^2\), the coefficients of \(\Lambda\) are proportional to \((c_j)\) since it kills \(bH^0(X,M)\). The radical of the diagonal form with weights \(c_j\) on \(W\) is \(\langle\mathbf1\rangle\): the diagonal form is nondegenerate on \(\mathbb C^m\) and the orthogonal complement of \(W\) is the line \(\langle\mathbf1\rangle\). Lifting gives the claimed kernel over \(\mathbb R\).

\textbf{Relation to the Hankel index.} For \(s\ge2\), the bundle \(\mathcal L^s\otimes T\) embeds \(X\) as an elliptic normal curve of degree \(m=3s\) in \(\mathbb P^{m-1}\). Green's syzygy bound \cite[Theorem 4.a.1]{cmp22} gives property \(N_{m-3}\), so the Hankel-index bound \cite[Corollary 2.7]{cmp3} gives rank at least \(m-2\) for every non-point extreme dual ray. Since \(g^s\) is strictly positive on the real curve, any extreme ray in its dual supporting face is non-point. Thus the rank lower bound also follows from that general theory. The proof below identifies the kernel and the unique Gram matrix and supporting ray at the specified \(g^s\), including \(s=1\).

\hypertarget{smooth-boundary-gram}{}\textbf{Uniqueness at a positive boundary point.} Let \(g\) be any positive definite quadratic on the plane. If \(g^s\) is on the boundary of \(\Sigma_s^T\), its reference Gram matrix is unique of rank two. Moreover, the supporting ray at \(g^s\) is unique, and the associated positive semidefinite bilinear form on \(U_s\) has rank \(3s-2\).

To prove this, put \(M=\mathcal L^s\otimes T\), of degree \(m=3s\). Identify \(M^2\) with \(\mathcal L^{2s}\) as above. Take a nonzero supporting functional \(\Lambda\) at \(w=g^s\) and write \(w=\sum_i a_i^2\). The positive semidefinite pairing \((v,v')\mapsto\Lambda(vv')\) kills every \(a_i\), so \(\Lambda\) vanishes on \(\sum_i a_i H^0(X,M)\). The span of the \(a_i\) has dimension at least two. For odd \(s\), a real square root would have a real zero since its divisor has odd degree \(3s\). For even \(s\), the divisor of a square root is \((s/2)\operatorname{div}(g)\) and hence has class \(\mathcal L^s\), which differs from \(M\) by the nontrivial \(T\).

Let \(B_0\) be the common zero divisor of the \(a_i\). It has no real support, since \(g^s\) is strictly positive on the real curve. If \(B_0\ne0\), choose two real linear combinations that form a pencil without base points after removing \(B_0\). Such a choice is obtained by fixing a nonzero section and choosing the second outside the finitely many proper subspaces given by the zeros of the first. Its evaluation sequence, tensored with \(M\), has kernel \(\mathcal O_X(B_0)\) and target \(M^2(-B_0)\). The kernel has vanishing \(H^1\), so \(\Lambda\) kills \(H^0(X,M^2(-B_0))\). The restriction \(H^0(X,M)\to H^0(B_0,M|_{B_0})\) is onto since \(M(-B_0)\) has positive degree, as does \(M^2(-B_0)\). Thus \(\Lambda\) descends to \(B_0\) and is nonnegative on all squares there. The element \(\iota^2=-1\) in the real algebra of this nonreal divisor, constructed in the strict-feasibility proof above, makes it vanish on all those squares. They span the target by polarization. This contradicts \(\Lambda\ne0\) and shows \(B_0=0\).

Choose two real linear combinations \(a,b\) of the \(a_i\) with no common zero. The pencil result gives the kernel and rank of the pairing and the one-dimensional annihilator.

Every positive semidefinite Gram matrix for \(w\) has range in that kernel, and \(a^2,ab,b^2\) are independent since \(a/b\) is nonconstant. The Gram matrix is therefore unique, and it has rank two by the exclusion of a single square. Any other supporting functional kills all products with its two Gram factors, so it lies on the same ray.

\Needspace{7\baselineskip}
\hypertarget{smooth-division-threshold}{}\textbf{General contraction paths.} Take any positive definite quadratic \(h\) with \(Q_1(h)\succ0\), and define \(\epsilon_s=\epsilon_s(h)\) for \(g_\epsilon=z^2+\epsilon h\). The monotonicity and strict-feasibility arguments above apply to this \(h\). Moreover, at a given level every \(\epsilon>\epsilon_s\) is strictly feasible: the expansion at \(\epsilon_s\) contains a positive multiple of \(h^s\), which has a positive definite Gram matrix, and every other term belongs to \(\Sigma_s^T\). Thus \(g_{\epsilon_s}^s\) lies on the boundary, with the unique rank-two Gram matrix just proved.

The threshold can be computed by a division condition on the conic intersection. For any \(\epsilon>0\), put \(Z_\epsilon=\operatorname{div}_X(g_\epsilon)\). It has degree six with no real support, since \(g_\epsilon\) is positive definite. Let \(D\) run over divisors obtained by taking exactly one point of each conjugate pair in \(Z_\epsilon\), with its whole multiplicity. There are at most four choices up to conjugation. Then \[
\epsilon_s=\min\left\{\epsilon>0:\ 
s\bigl([D]-[\mathcal L]\bigr)=T\ \text{in }\operatorname{Pic}^0(X_{\mathbb C})
\text{ for one such }D\right\}.
\] Indeed, at the threshold the two real Gram factors \(a,b\) have no common zero. The sections \(a+ib\) and \(a-ib\) then have disjoint divisors summing to \(sZ_\epsilon\), so their divisors are \(sD,s\overline D\), respectively. Conversely the class equality gives a complex section of \(M\) with divisor \(sD\). Its product with its conjugate is a positive constant times \(g_\epsilon^s\), as is seen by evaluating at a real point. Its real and imaginary parts give feasibility. This proves the minimum characterization. With the flex \(O\) as the group origin, the equality says \([s](\sum_{P\in D}P)=P_0\), counting multiplicities \cite[Proposition III.3.4]{cmp20}. This gives an algebraic computation with the six conic intersections and an elliptic division point for algebraic input metrics.

\textbf{Analytic dependence on the metric.} For each fixed \(s\), the threshold and critical Gram matrix are real analytic on the open set of positive definite \(h\) with \(Q_1(h)\succ0\). Let \(g=g_{\epsilon_s}\) and let \(\Lambda\) span its supporting ray. Since \(h\) is a sum of squares in \(U_1\), the binomial expansion of \(g^{s-1}h\) shows that every term belongs to \(\Sigma_s^T\). The term \(\epsilon_s^{s-1}h^s\) is strictly feasible, so \(\Lambda(g^{s-1}h)>0\).

Write \(g^s=a^2+b^2\) at the threshold. The derivative of the map \((a,b)\mapsto a^2+b^2\) has image the hyperplane annihilated by \(\Lambda\) and kernel spanned by \((b,-a)\). Restrict to an affine slice through \((a,b)\) transverse to this kernel and add \(\epsilon\) as an unknown in \(a^2+b^2=g_\epsilon^s\). The positive pairing just proved makes the derivative in these unknowns invertible. The implicit function theorem gives analytic factors and a parameter for nearby \(h\). Their pencil remains without base points, an open condition, so its product image is still a hyperplane. Choose its annihilator continuously with the original sign. The associated pairing has the moving pencil in its kernel and stays positive definite on a complementary subspace, by the rank statement at the original metric. It is therefore still positive semidefinite. Thus the analytic solution is on the boundary and gives the threshold and its unique Gram matrix. Differentiating the equality yields, for a quadratic perturbation \(k\), \[
D\epsilon_s(h)[k]=-\epsilon_s\,
\frac{\Lambda(g^{s-1}k)}{\Lambda(g^{s-1}h)}.
\] In particular, at a reflection-invariant \(h\) the first variation vanishes for a perturbation odd under \(y\mapsto-y\), since simultaneous reflection preserves the test and changes the sign of the perturbation.

\textbf{A rational example with odd terms.} On \(y^2z=x^3+x^2z+xz^2\), take \[
h=\tfrac{37}{36}(x^2+z^2)+y^2+xz+\tfrac13xy-\tfrac43yz .
\] Both \(h\) and \(Q_1(h)\) are positive definite. The level-two threshold is exactly \(\epsilon_2=1/48\). To see this, put \(G=h+48z^2\) and work on \(z=1\). With \(F=(x+1)(x-4)+(\tfrac16+3i)(y+i)\), the identities \(F\overline F=(x+1)G\) and \((x+iy)(x-iy)=x(x+1)^2\) show that \(F^2(x-iy)/(x+1)^2\) has squared modulus \(xG^2\). Its real and imaginary parts are regular polynomials in the six reference basis elements. The supplement verifies the identity and a supporting functional by exact rational arithmetic: its matrix kills the two displayed factors and has a positive definite principal submatrix of order four. Thus \(G^2\), in the twisted-square criterion, lies on the boundary. Strict feasibility for larger parameters identifies the asserted threshold.

\Needspace{12\baselineskip}
\hypertarget{smooth-cubic-sixth-power}{}\textbf{Quartic thresholds along reflection-invariant paths.} In the Weierstrass coordinates above let \(p_2(x)=x^2+Ax+B\), with \(0\) the smallest real root of \(xp_2(x)\). Let \(h=a x^2+b y^2+c z^2+2e xz\) be positive definite with \(Q_1(h)\succ0\), and let \(\epsilon_s\) be the exactness threshold for \(g_\epsilon=z^2+\epsilon h\). Write \[
L=\int_0^\infty\frac{dr}{\sqrt{r p_2(-r)}} .
\] At every level \(s\ge1\), the threshold is explicit and its reference Gram matrix is unique of rank two. For \(s\ge2\) the formula uses one elliptic division point and the distinguished real root of a quartic. As \(s\to\infty\), \[
\epsilon_s=\frac{L^6}{4096b\,s^6}
\left(1-\frac{a L^2}{32b\,s^2}+O(s^{-4})\right),\qquad
\lim_{\epsilon\downarrow0}\epsilon^{1/6}s_*(\epsilon)=\frac{L}{4b^{1/6}},
\] where \(s_*(\epsilon)\) denotes the first exact level. These conclusions hold for every \(d\ge3\).

\emph{The threshold at level one.} Put \(\widehat a=a/b\), \(\alpha=A+\widehat a\), \(\beta=B+2e/b\) and \(C=(c+1/\epsilon)/b\). The level-one matrix, after removing the scalar \(b\epsilon\), has a scalar block \(C/B\) and a two-by-two block of diagonal entries \(1,\beta-AC/B\) and off-diagonal entry \((\alpha-C/B)/2\). Completing the square in its determinant gives \[
C_1=B\left(\widehat a-A+2\sqrt{\beta-A\widehat a}\right),\qquad
\epsilon_1=\frac1{bC_1-c}.
\] Indeed, as \(\epsilon\) decreases from infinity, \(C/B\) increases from \(c/(bB)\), which lies strictly inside the determinant interval by \(Q_1(h)\succ0\). At its upper endpoint the unique matrix has rank two.

\emph{The degree-four norm construction for \(s\ge2\).} Removing the positive scalar \(b\epsilon\) gives the criterion \[
G_C^s=xU+p_2V,\qquad G_C(x)=x^3+\alpha x^2+\beta x+C,
\] where \(U,V\) are globally nonnegative and have degrees at most \(2r_s,2t_s\). Work on the auxiliary curve \(w^2=-xp_2(x)\), with point at infinity \(O\) and \(P_{\mathrm{aux}}=(0,0)\). Define \(q=q_s>0\) by \[
\int_q^\infty\frac{dr}{\sqrt{r p_2(-r)}}=\frac{L}{s}.
\] Substitution \(r\mapsto B/r\) interchanges the two halves of this integral at \(\sqrt B\), so \(q\ge\sqrt B\). Also \(A<2\sqrt B\), since \(p_2(-r)>0\) for every \(r>0\). The differential \(dx/w\) gives an additive coordinate \(\xi\) on the real component of \(O\), with period \(2L\) \cite[Propositions III.5.1 and VI.5.2]{cmp20}. On the branch \(w>0\), \(P_{\mathrm{aux}}\) has coordinate \(L\) and \(P=(-q,\sqrt{q p_2(-q)})\) has coordinate \(p=L/s\). Hence \(sP=P_{\mathrm{aux}}\). Equivalently, \(q_s\) is the largest positive \(q\) on this branch for which \(sP=P_{\mathrm{aux}}\). By the identification with \(\operatorname{Pic}^0\), choose a real function \(k\) with divisor \(sP+P_{\mathrm{aux}}-(s+1)O\) \cite[Proposition III.3.4]{cmp20}. Its norm \(k\bar k\), with the bar denoting \(w\mapsto-w\), is a constant times \(x(x+q)^s\). The constant is positive by evaluation where \(x\) is large positive and \(w\) is purely imaginary. Scale \(k\) to make this constant one.

Set \[
\rho=\sqrt{q-A+B/q},\qquad S=\sqrt{\widehat a-A+3q}.
\] Let \(\Delta\) be the unique real root greater than \(S\) of \[
\Phi(t)=\frac{t^4}{4}-\frac{S^2t^2}{2}-2q\rho t
+\frac{(\alpha+q)^2}{4}-q^2-\beta=0.
\] To see existence, the determinant condition at \(h\), with \(\omega=c/(bB)\), gives \[
\beta>A\omega+\tfrac14(\alpha-\omega)^2
=A\widehat a+\tfrac14\bigl(\omega-(\widehat a-A)\bigr)^2
\ge A\widehat a+\tfrac14(A-\widehat a)_+^2 ,
\] where \(t_+=\max(t,0)\). Therefore \[
\begin{aligned}
\Phi(S)
&=A\widehat a-\beta-q\widehat a+2Aq-3q^2-2q\rho S\\
&<2Aq-3q^2-q\widehat a-\tfrac14(A-\widehat a)_+^2\\
&\le 2Aq-3q^2-\tfrac14(A_+)^2<0 .
\end{aligned}
\] Here \(-qx-\tfrac14(A-x)_+^2\) is decreasing for \(x\ge0\) because \(A<2q\). The last inequality is immediate for \(A\le0\), and for \(0<A<2q\) the expression is \(-(A-2q)(A-6q)/4\). The polynomial tends to \(+\infty\) and has strictly increasing derivative on \([S,\infty)\), giving the unique root there.

Define \[
\ell=\frac{\alpha+q-\Delta^2}{2},\qquad
\gamma=-q(q-\ell+\rho\Delta),\qquad
C=\frac{\gamma^2}{q},\qquad
f=x^2+\ell x+\gamma+\Delta w .
\] These definitions give \(f(P)=0\), since \(w(P)=q\rho\). The norm \(f\bar f\) agrees with \((x+q)G_C\) in the leading, cubic and constant coefficients. Equality of the linear coefficients is equivalent to \(\Phi(\Delta)=0\). Both polynomials vanish at \(x=-q\), so the quadratic coefficients agree as well. We have \(\ell<A-q\) and \(\gamma<0\), using \(A<2q\). The resulting norm identity gives the required two squares. Indeed, \(f^s\bar k/(x+q)^s\) is regular on the affine curve by vanishing to order at least \(s\) at both \(P\) and \(-P\). At \(O\) its pole order is at most \(3s+1\), and it vanishes at \(P_{\mathrm{aux}}\). The pole orders of \(x,w\) give \[
xu+wv=\frac{f^s\bar k}{(x+q)^s},\qquad
\deg u\le r_s,\quad \deg v\le t_s .
\] Taking the norm and dropping a factor \(x\) gives \(G_C^s=xu^2+p_2v^2\).

\emph{The norm factors have no common zero.} We use the same construction on the connected parameter domain
\[
B>0,\qquad A<2\sqrt B,\qquad \widehat a>0,\qquad
\beta>A\widehat a+\tfrac14(A-\widehat a)_+^2.
\]
These conditions suffice for the preceding inequalities that define \(\Delta>S\), \(\ell<A-q\) and \(\gamma<0\). The domain is connected since the first three inequalities define a connected base and the last is a strict epigraph over it. Allow the split nodal fiber \(A=-2\sqrt B\). The integral defining \(q\) remains regular there, and the auxiliary curve has normalization \(x=-t^2\), \(w=t(t^2+\sqrt B)\). With \(Q=\sqrt q=B^{1/4}\cot(\pi/(2s))\), take \(k=t(t-Q)^s\). At the two node preimages \(t=\pm i B^{1/4}\) its values agree, since their ratio is \(-\exp(-2is\arctan(B^{1/4}/Q))=1\). Thus it is a polynomial in \(x,w\): writing a polynomial as an even part plus \(t\) times an even part, equality at the preimages makes that latter coefficient divisible by \(t^2+\sqrt B\). Its norm is \(x(x+q)^s\) and its zeros are \(P\) with multiplicity \(s\) and \(P_{\mathrm{aux}}\) with multiplicity one.

The coefficients of \(k\) can be chosen locally continuously on this domain. In the fixed monomial basis of functions of pole order at most \(s+1\) at \(O\), its vanishing at \(P_{\mathrm{aux}}\) and to order \(s\) at \(P\) is a linear system of constant nullity one. For a smooth cubic this follows from the divisor of \(k\). On the nodal normalization it follows because a polynomial of degree at most \(s+1\) with these zeros is a multiple of \(t(t-Q)^s\). The constraints vary continuously since the integral for \(q\) does and \(x\) is a local parameter at \(P\). The positive norm normalization can be imposed continuously locally. Division by the monic \((x+q)^s\) in the two normal-form coefficients then makes \(u,v\) locally continuous.

On the original cubic, homogenize \(xu\) and \(yv\) at degree \(s+1\) and remove the fixed divisor \(P_0+2O\). These are sections of \(M=\mathcal L^{s+1}(-P_0-2O)\), with \(M^2\simeq\mathcal L^{2s}\) via \(xz\), and their squares add to \(g_C^s\), where \(g_C=\widehat a x^2+y^2+(\beta-B)xz+Cz^2\). The sections have no common zero, even on the nodal fiber. At \(P_0,O\), use the identity after removal of the fixed divisor and the nonzero values \(g_C(P_0)=C\), \(g_C(O)=1\). Elsewhere work on the auxiliary curve, using \(y=iw\). At a point away from \(P,-P,P_{\mathrm{aux}},O\), a common zero would require \(f=\bar f=0\). This forces \(w=0\) and \(x^2+\ell x+\gamma=0\). At \(x=0\) the latter is impossible. At a root of \(p_2\) it reduces to \((\ell-A)x+\gamma-B\), which can vanish only at a negative real \(x\) since both coefficients are negative. There are no negative real roots of \(p_2\). This also excludes the node. At \(-P\), \(f(-P)=-2\Delta q\rho\ne0\), and the order-\(s\) zero of \(\bar k\) cancels the denominator exactly. Thus the norm factor is nonzero at \(-P\), and its conjugate under \(w\mapsto-w\) is nonzero at \(P\).

\emph{A positive dual form at a split node.} Set \(A=-2\), \(B=1\) and choose
\[
R=\cot\bigl(\pi/(6s)\bigr),\qquad Q=\cot\bigl(\pi/(2s)\bigr),\qquad
\widehat a=2+3R^2,\qquad \beta=3R^4.
\]
On the auxiliary normalization \(w=t(t^2+1)\), the polynomial \((t-Q)(t+R)^3\) has the required form \(x^2+\ell x+\gamma+\Delta w\). Indeed \(Q=(R^3-3R)/(3R^2-1)\) makes its linear and cubic coefficients equal, \(\gamma=-QR^3\), and \(\Delta=3R-Q\). Since \(0<Q<R/3\) and \(R>\sqrt3\), one has \(\Delta^2-S^2=6R^2-6RQ-2Q^2-4>0\). Its norm is \((x+q)(x+R^2)^3\), so the distinguished root gives \(C=R^6\).

On the original nodal cubic use the normalization \((t,u)\mapsto(t^2u,t(t^2-u^2),u^3)\). Its pullbacks of \(g_C\) and \(xz\) are \((t^2+R^2u^2)^3\) and \((tu^2)^2\). Any SOS for \(g_C^s xz\) pulls back to summands divisible by \(tu^2\), by their vanishing orders at its two real zeros. After dividing, their degree is \(3s\) and their vectors at \((1,1)\) and \((-1,1)\) are opposite, since these parameters map to the same point with the same homogeneous representative. The endpoint angle in the metric \(t^2+R^2u^2\) is \(2\arctan(1/R)=\pi/(3s)\). If \(R\) is increased, using the same coefficient formulas, the angle becomes less than \(\pi/(3s)\) and the \hyperlink{edge-endpoint-bound}{binary endpoint bound} rules this out. A degree-\(3s\) unit vector on the circle can move by a spherical distance of at most \(3s\) times the endpoint angle, whereas the opposite vectors are at distance \(\pi\). At the displayed \(R\) the norm construction provides an SOS. Consequently \(g_C^s\) lies on the boundary of the closed square cone for \(M\). Closedness follows by the same finite real evaluation argument as for the smooth curve. A nonzero supporting functional has a positive semidefinite pairing and kills products with the two norm factors.

\hypertarget{smooth-signature}{}\textbf{Lemma 3.2 (Dual signature in fixed coordinates).} Fix \(s\ge2\). Consider a connected parameter domain of the above plane cubics, with smooth and geometrically integral split nodal fibers allowed. Suppose two sections of \(M=\mathcal L^{s+1}(-P_0-2O)\) vary locally continuously in the affine coordinates below and have no common complex zero in each fiber. The line of functionals annihilating their products with \(H^0(X,M)\) has a positive semidefinite generator throughout the domain if it has one at a single parameter.

To specify the common coordinates, take the pairs \((u,v)\) with \(\deg u\le r_s\) and \(\deg v\le t_s\), representing the section \(xu+yv\) on \(z=1\) before removing the fixed divisor \(P_0+2O\). This space has dimension \(3s\). The same pole-order description holds on the nodal fiber, since \(P_0,O\) remain smooth and the affine coordinate ring has basis \(1,y\) over \(\mathbb R[x]\). The product target is the \(6s\)-dimensional space of \(H+yJ\) with \(\deg H\le3s\) and \(\deg J\le3s-2\). Division by \(xz\) on homogeneous products gives the multiplication map in the affine chart,
\[
\mu_{A,B}\big((u,v),(u',v')\big)=xuu'+y(uv'+vu')+p_2vv'.
\]
Its matrix in the fixed monomial bases depends polynomially on \(A,B\). For a pair of sections \(a_1,a_2\) as in the lemma, the map \((b_1,b_2)\mapsto a_1 b_1+a_2 b_2\) has rank \(6s-1\) by the base-point-free pencil calculation. Its matrix admits a nonzero minor of size \(6s-1\) near each parameter, so the annihilator line has a local continuous nonzero generator \(\Lambda\). The symmetric pairing \(\Lambda\circ\mu_{A,B}\) has rank \(3s-2\) by the same pencil calculation. On a local choice of generator the nonzero eigenvalues keep their signs. Hence the property that one choice of sign yields a positive semidefinite pairing is both open and closed, and connectedness proves the assertion.

Apply \hyperlink{smooth-signature}{Lemma~3.2} to the two norm factors. The split nodal parameter provides the positive semidefinite generator, so at every smooth parameter a nonzero positive dual functional annihilates \(g_C^s\). The argument takes place in the square cone of sections on the parameter domain above. The physical condition for membership on the admissible metric path will follow from \(bC>c\).

We first check that \(bC>c\). Fix the curve and level, and rescale the metrics to \(b=1\). The set of admissible \((a,c,e)\) is convex, since positive definiteness of \(h\) and \(Q_1(h)\) is defined by linear matrix inequalities. The root \(\Delta>S\) is simple and varies continuously, so \(C-c\) is continuous. It cannot vanish: at \(C=c\), the supporting functional would annihilate \(h^s\), which has a positive definite Gram matrix by strict feasibility after one level. There is a metric with \(C>c\). Choose \(e=0\) and \(a>0\) with \(|A+a|<2\sqrt B\), possible because \(A<2\sqrt B\). For these coefficients \(Q_1(h)\succ0\) holds for all small positive \(c\), and \(C>0\) is independent of \(c\). This proves \(bC>c\) throughout the admissible set.

Set \(\widehat\epsilon_s=1/(bC-c)\). Then \(g_{\widehat\epsilon_s}=b\widehat\epsilon_s g_C\). Its power is feasible and is annihilated by a nonzero positive dual functional. Along the admissible path the only feasible boundary point at this level is \(\epsilon_s\), by the strict feasibility above the threshold. Thus \(\epsilon_s=\widehat\epsilon_s\).

\emph{Asymptotic coefficients.} Put \(p=L/s\). The integral tail is \(2q^{-1/2}+(A/3)q^{-3/2}+O(q^{-5/2})\), so \(q=4p^{-2}+A/3+O(p^2)\). The limit of \(4\Phi(\lambda\sqrt q)/q^2\) is \((\lambda+1)^3(\lambda-3)\), uniformly on bounded intervals. Its signs at \(2,4\) bracket the root for large \(q\), so \(\Delta/\sqrt q\to3\) and \(\gamma/q^2\to-8\). To expand this branch, compare coefficients in the norm identity: \[
\ell=\frac{\gamma^2/q-B\alpha+q(\beta-B)}{2(\gamma-B)},\qquad
\ell^2-2A\ell+2\gamma=\beta+\alpha q-A(\alpha+q).
\] For \(\kappa=-\gamma/q^2\), the second equation after division by \(q^2\) has limiting equation \(\kappa^2/4-2\kappa=0\) with a simple root at \(8\). Its analytic branch has \[
\gamma=-8q^2+\frac{9A-\alpha}{2}q+O(1),
\] by comparing the coefficient of \(q\) in the equation for \(\ell,\gamma\). Substitution yields \(C=4096p^{-6}+128(a/b)p^{-4}+O(p^{-2})\). Expanding \(1/(bC-c)\) gives the stated two terms. The first exact level limit follows by sandwiching \(\epsilon\) between adjacent thresholds.

\textbf{Two exact examples on one smooth cubic.} Consider \(X\) with \[
F=y^2z-x^3-xz^2=0,\qquad O=[0:1:0],\qquad P=[0:0:1].
\] The Weierstrass cubic \(x^3+x\) has simple roots and just one real root, so \(X\) is smooth and \(P-O\) gives its only nontrivial real two-torsion class. We use \(f_T=xz\). Its divisor is \(2P+4O\): in the chart \(z=1\) at \(P\), \(x\) has order two, and at \(O\) in the chart \(y=1\), \(x\) has order one and \(z\) has order three. The class of the half divisor relative to the hyperplane class \(3O\) is \(P-O\). The form \(xz\) is nonnegative on \(X(\mathbb R)\), since \(x/z\ge0\) at every real point with \(z\ne0\). It is not SOS at level zero, since a nonzero linear summand would have to have divisor \(P+2O\), which represents the nontrivial class relative to \(3O\).

For \(g=a x^2+b y^2+c z^2\), with \(a,b,c>0\), the matrix in \hyperlink{res-1F}{Theorem~1F} has diagonal \((b,b,c)\) and sole possibly nonzero off-diagonal entry \((a-c)/2\) in the upper left block. Thus level one is exact precisely when \(|a-c|\le2b\). For \(a=b=c=1\) the identity reduces to \[
(x^2+y^2+z^2)xz=(x^2)^2+(xz)^2+(yz)^2\pmod F .
\]

Now take \((a,b,c)=(13/4,1,1)\). Level one fails because the unique Gram matrix has principal minor \(-17/64\). At level two we use a basis of \(V_{T,2}\) grouped into two blocks. \[
e=(x^3,x^2z,xz^2)^{\mathsf T},
\qquad
o=(x^2y,xyz,yz^2)^{\mathsf T}.
\] Expansion modulo \(F\) gives \[
g^2 xz=e^{\mathsf T}
\begin{pmatrix}
2ab&a^2/2&0\\
a^2/2&2b(a+c)&(2ac-c^2)/2\\
0&(2ac-c^2)/2&2bc
\end{pmatrix}e
+o^{\mathsf T}\operatorname{diag}(b^2,b^2,c^2)o .
\] At the chosen metric the leading principal minors of the first block are \(13/2\), \(28015/1024\) and \(2847/512\). Both blocks are positive definite, so level two is exact. These first levels hold for all \(d\ge3\) by the reference test.

\Needspace{9\baselineskip}
\hypertarget{smooth-exact-all-levels}{}\textbf{All thresholds on \(y^2z=x^3+xz^2\).} For \(g_\epsilon=z^2+\epsilon(x^2+y^2+z^2)\), level one has threshold \(\epsilon_1=1/2\). At each \(s\ge2\) use \(q_s\) and \(L\) above with \(p_2(x)=x^2+1\). Let \(\gamma_s\) be the unique root with \(\gamma_s<-8q_s^2\) of \[
(\gamma^2-q_s)^2+4q_s^2(\gamma-1)^2(2\gamma-q_s-1)=0 .
\] Then \(\epsilon_s=(\gamma_s^2/q_s-1)^{-1}\), and its reference Gram matrix is unique of rank two.

Indeed, substitution \(r\mapsto1/r\) in the integral gives \(q_2=1\), hence \(q_s\ge1\). At \(\gamma=-8q_s^2\) the first quartic term is less than \(4096q_s^8\) and its second term is less than \(-4096q_s^8\). A root \(\gamma<-8q_s^2\) therefore exists. With \(C=\gamma^2/q_s\), \(\ell=(C-1)/(2(\gamma-1))\), and \(\Delta=\sqrt{1+q_s-2\ell}\), the quartic gives the norm identity and \(f\) vanishes at \(P\) for every \(s\ge2\), since \(\ell<0\) and \(|\ell|q_s<|\gamma|/2\), so \(q_s^2-\ell q_s+\gamma<0\). Since \(-\gamma>8q_s^2\) and \(q_s\ge1\), the displayed formula for \(\ell\) gives \(-\ell>q_s\). Hence \(\Delta>S\), so this solution is on the distinguished quartic branch. The definitions then determine \(\gamma\) uniquely.

For illustration \(q_2=1\) gives \(\gamma_2=-(5+4\sqrt2)\) and \(\epsilon_2=(5\sqrt2-7)/8\). At level three, \(q_3=\sqrt{3+2\sqrt3}\). In fact for this value \(q\) the function \(x^2+(3q-q^3)x/2-\sqrt{q^3}\,w\) has norm \(x(x+q)^3\), by \(q^4-6q^2-3=0\). It vanishes at \(P_{\mathrm{aux}}\) and to order three at \(P=(-q,\sqrt{q^3+q})\) on the branch \(w>0\). Hence \(3P=P_{\mathrm{aux}}\) and \(q=q_3\). The quartic gives \(0.0008290238<\epsilon_3<0.0008290239\). The supplement includes separate primal and dual checks at these two levels.

\textbf{A one-component example at level two.} For \(A=B=1\) take \(h=x^2+y^2+xz+z^2\). This admissible metric gives \(q_2=1\), \(\Delta=3\), and \(\epsilon_2=1/48\). In the fixed coordinates of \hyperlink{smooth-signature}{Lemma~3.2} the norm factors give
\[
(x^3+2x^2+2x+49)^2
=x(7x^2-41x+7)^2+(x^2+x+1)(-x^2+23x-49)^2.
\]
The accompanying checks verify both rank-two dual blocks in this example and in the split-real example below.

\textbf{A split real example at level two.} Take \(A=-7\), \(B=1\), and \(h=6x^2+y^2+z^2-2xz\). Both \(h\) and \(Q_1(h)\) are positive definite. Here \(q_2=1\), \(\rho=3\), \(S^2=16\), and the quartic has \(\Delta=6\) as its root above \(S\). This gives \(\ell=-18\), \(\gamma=-37\), \(C=1369\), and \(\epsilon_2=1/1368\). Explicitly, \[
(x^3-x^2-x+1369)^2
=x(15x^2-546x+3663)^2
+(x^2-7x+1)(-x^2+110x-1369)^2 .
\] The roots of the first square factor lie in \((8,9)\) and \((27,28)\), and those of the second lie in \((14,15)\) and \((95,96)\), by sign changes at the indicated endpoints. All four lie above the largest root of \(xp_2(x)\), which is less than \(7\). These brackets give an independent check of the locations of the roots in this example.

\subsection{3.4 Exact levels for a line and a conic}\label{s-3-4}

\begin{paperresult}{Theorem 1G}{res-1G}
Let \(X\) be the union of a real projective line and a smooth real conic with real points, and let \(g\) be positive definite. Suppose first that the components intersect transversely in two real points. Parametrize the conic as \[
\nu(t,u)=Ut^2+Wtu+Vu^2,
\] where \(g(U)=g(V)=1\) and \([U],[V]\) are the intersection points. Put \(G=g\circ\nu\), \(\gamma=\langle U,V\rangle_g\), \(\phi=\arccos|\gamma|\), and \(\sigma=\operatorname{sign}(\gamma)\) with \(\sigma=1\) at \(\gamma=0\). Let \(m_s=(t^{2s},t^{2s-1}u,\ldots,u^{2s})^{\mathsf T}\). For every \(d\ge3\) and \(s\ge1\), the level-\(s\) PPT hierarchy on \(X\) is exact if and only if there is a real matrix \(B\) satisfying \[
B\succeq0,\qquad m_s^{\mathsf T}B m_s=G^s,\qquad B_{1,2s+1}\in-\sigma^s J_s(\phi),
\] with \(J_s\) as in \hyperlink{res-1C}{Theorem~1C}. In particular, for \(\alpha=\langle U,W\rangle_g\), \(\beta=\langle V,W\rangle_g\) and \(\kappa=g(W)\), level one is exact exactly when \[
\begin{pmatrix}
1&\alpha&-\gamma\\
\alpha&\kappa+4\gamma&\beta\\
-\gamma&\beta&1
\end{pmatrix}\succeq0.
\] Level zero is inexact. The first exact level is independent of \(d\) and is at most \(\lceil\pi/\phi\rceil\). On each fixed such \(X\), it has minimum one over positive definite \(g\) and is unbounded over those \(g\).

If the two transverse intersections are complex conjugates, the first exact level is one when both lie on \(V(g)\) and no level is exact otherwise. If the line is tangent to the conic, no level is exact.

\end{paperresult}

Let \(X=L\cup C\) with the components as in \hyperlink{res-1G}{Theorem~1G}. For transverse intersections, a section of degree \(j\ge1\) is a pair of binary forms of degrees \(j\) on \(L\) and \(2j\) on \(C\), agreeing at the two nodes in the same homogeneous representatives. Pullback is injective. On the two components the pair has \(3j+2\) coefficients and the two gluing equations are independent, using evaluation on the line. For conjugate nodes this counts two real equations. Thus the space of pairs has dimension \(3j\), the dimension of \(R_j\), proving the identification.

\textbf{Two real nodes.} Use \((t,u)\mapsto Ut+Vu\) on the line and \(\nu\) on the conic. In the degree-\(2d\) cone, each nonzero component of an extreme ray is a square with only real roots in its square root. Indeed, after extracting its real zeros as a square, a positive-degree residual strictly positive binary factor can be perturbed by a form of the same degree vanishing at both endpoints, preserving the two gluing equations. If the other component is nonzero the perturbation is necessarily nonproportional. If that component is zero, the strictly positive factor is nonzero at the endpoints, so the perturbation again cannot be proportional. The compact-base argument of \hyperref[s-3]{Section~3} then reduces the SOS inclusion to these extreme rays.

For a nonnegative pair with square roots \(p_L,p_C\) and nonzero node values, rescale a vector of SOS summands on each component by dividing by its square root. The resulting vectors have squared norms \((g|_L)^s\) and \(G^s\). There are two possible relative sign patterns between the two square roots at the nodes, up to changing one square root by a global sign. When they agree at both nodes the roots already descend, and multiplication by \(g^s\) preserves SOS. In the other case choose the roots to agree at the first node and have opposite signs at the second. The quotient vectors can be glued after multiplication by the roots exactly when their endpoint inner products are negatives of each other. The endpoint norms are one for both quotient vectors. The attainable inner products on the line form \(\sigma^s J_s(\phi)\) by \hyperref[s-3]{Section~3}. On the conic they are precisely the entries \(B_{1,2s+1}\) of matrices with \(B\succeq0\) and \(m_s^{\mathsf T}B m_s=G^s\), by Gram factorization. For inner products with opposite signs, the original vectors obtained by multiplying back by their roots have the same two-by-two endpoint Gram matrix. Pad them with zeros and apply an orthogonal transformation on one component to match both endpoints. Their entries then descend to \(X\). When one endpoint is zero, any quotient SOS on each nonzero component gives matching endpoint Gram matrices after multiplication by the roots. When one component is zero the other is handled by extending by zero, since both endpoint values vanish.

Conversely, choose \(p_L\) and \(p_C\) with distinct real roots away from the endpoints, with values \((1,1)\) on the line and \((1,-1)\) on the conic. For either binary degree, a product of linear factors in endpoint coordinates can attain the prescribed values by choosing nonzero distinct roots of prescribed product. The squares give a nonnegative form on \(X\). All summands in an SOS for its product by \(g^s\) restrict to forms divisible by the corresponding roots, since all their roots are real and simple. Their quotients therefore have opposite endpoint inner products, proving the stated Gram condition. This also proves inexactness at level zero, since constant quotient vectors have inner product one on each component. \hyperlink{support-prop}{Proposition~3.1} identifies the SOS criterion with exactness of the PPT hierarchy.

At \(s=1\) the line correlation is \(\gamma\). In the basis \((t^2,tu,u^2)\) the conic quartic \(G\) has coefficients \(1,2\alpha,\kappa+2\gamma,2\beta,1\). The required endpoint entry \(-\gamma\) fixes the whole Gram matrix, giving the displayed condition. For \(s\ge\lceil\pi/\phi\rceil\), the interval \(J_s(\phi)\) is all of \([-1,1]\). A Gram representation of \(G^s\) exists by binary factorization and its endpoint entry automatically lies in this interval.

\textbf{Examples and variation on a fixed curve.} Take \(C:xz=y^2\), \(L:y=0\), and \(\nu(t,u)=(t^2,tu,u^2)\). The metric \(x^2+y^2+z^2\) makes \(U,W,V\) orthogonal and gives exactness at level one. For a first exact level of two, use \(x^2+y^2+z^2-xz\), so \(\gamma=-1/2\), \(\kappa=1\) and \(\alpha=\beta=0\). The middle diagonal entry at level one is \(-1\). At level two the conic square \((t^4+u^4)^2\) has the diagonal Gram matrix \(\operatorname{diag}(1,0,2,0,1)\) in degree four, and its endpoint correlation is zero. On the line \((t^2-tu+u^2)^2\) has a positive definite Gram matrix with the same endpoint correlation, \[
\begin{pmatrix}1&-1&0\\-1&3&-1\\0&-1&1\end{pmatrix}.
\] Thus level two is exact.

\textbf{An angle formula when the conic metric is a square.} Suppose in addition that \(G=q^2\), where \(q=t^2+2btu+u^2\) is a positive definite binary quadratic, with \(\psi=\arccos|b|\). The attainable conic correlations at level \(s\ge1\) are \(J_{2s}(\psi)\) by \hyperref[s-3]{Section~3}, since its squared norm is \(q^{2s}\). The sign change of a unit lift in the \(q\) metric does not affect the conic correlation because the binary degree is even. If \(\gamma\ge0\), the line correlations are \(J_s(\phi)\). For \(s\ge2\), both intervals have right endpoint one, so a correlation and its negative can be matched exactly when their two left endpoints sum to at most zero. At \(s=1\) the line value is \(\cos\phi\), and the same sum test applies using that singleton. Using \(\cos a+\cos b\le0\) for \(a,b\in[0,\pi]\) exactly when \(a+b\ge\pi\), we obtain exactness if and only if \[
s(\phi+2\psi)\ge\pi .
\] Clipping either angle at \(\pi\) does not change this inequality.

To include \(\gamma<0\), write the matrix of \(g\) in the basis \((U,W,V)\). Comparing \(G\) with \(q^2\) gives \[
M(\gamma,b)=
\begin{pmatrix}
1&2b&\gamma\\
2b&2(1-\gamma)+4b^2&2b\\
\gamma&2b&1
\end{pmatrix}.
\] Its determinant is \(2(1-\gamma)^2(1+\gamma-2b^2)\). Positive definiteness therefore implies \(2b^2-1<\gamma\). Thus \(\cos(2\psi)<\cos(\pi-\phi)\) and \(\phi+2\psi>\pi\). Moreover, \(J_2(\psi)\) contains \(-\gamma\), the conic value needed to match the negative of the single line value \(\gamma\) at level one. Exactness at level one implies exactness at every higher level by multiplication by the SOS \(g\). Consequently the same angle formula holds for either sign of \(\gamma\).

On any fixed line and conic with two real nodes, choose \(g_c\) with matrix \(M(c,c)\) in the fixed basis \((U,W,V)\), \(0\le c<1\). This matrix is positive definite: its symmetric block on the sum of the endpoint coordinates and the middle coordinate has top-left entry \(1+c\) and determinant \(2(1-c)(1+c-2c^2)>0\), and the antisymmetric direction has eigenvalue \(1-c\). The pullback \(G=q_c^2\) with \(q_c=t^2+2ctu+u^2\) has \(\phi=\psi=\arccos c\). This proves the common formula of \hyperlink{res-1H}{Corollary~1H} for a line and conic with two real nodes.

\textbf{Two conjugate nodes.} Choose conjugate lifts for the nodes and compatible representatives on both components. On a component \(i\), write \(q_i\) for a positive definite real binary quadratic vanishing at the two node parameters. On an extreme ray nonzero on both components, each restriction is a square with only real roots in its square root. The same perturbation proof works with a real perturbation divisible by \(q_i\) on a component whose strictly positive residual factor has degree at least two. The perturbation is nonproportional as the other component is nonzero. At a node the two roots have equal or opposite complex values because their squares agree. The sign is the same at its conjugate, so a single sign change suffices to make the roots descend. If the extreme ray is supported on just one component \(i\), its nonnegative restriction must be divisible by \(q_i\). The component-supported face is identified by division by \(q_i\) with the cone of nonnegative binary forms of degree two less. Its extreme rays are therefore \(q_i p_i^2\) with \(p_i\) having only real roots, by the same factorization argument without gluing constraints.

Suppose the nodes lie on \(V(g)\). At level one a two-component extreme ray that is a global square remains a sum of squares after multiplication by \(g\). On a ray supported on component \(i\), the product restricts to \(q_i^2(g_i/q_i)p_i^2\), where \(g_i\) is the pullback of \(g\) and \(g_i/q_i\) is nonnegative. Binary factorization gives an SOS with every summand divisible by \(q_i\), hence extendible by zero to the other component. Thus level one is exact. For level zero, a supported ray \(q_i p_i^2\) is not SOS. Every proposed summand is zero on the other component and hence divisible by \(q_i\) on component \(i\). The restriction has only a simple zero at either nonreal node, since the roots of \(p_i\) are real. Finally, if \(g\) does not vanish at the nodes, the same simple-zero witness persists at every level. In the tangent case \hyperlink{res-1D}{Theorem~1D} gives inexactness at every level.

For example, take \(C:x^2+y^2=z^2\) and \(L:z=0\). The form \(g=x^2+y^2+z^2\) vanishes at the conjugate nodes, and the first exact level is one. The binary quadratic \(t^2+u^2\) cuts out the nodes both on the line parametrized by \((t,u,0)\) and on the conic parametrized by \((t^2-u^2,2tu,t^2+u^2)\). The restrictions of \(g\) are \(t^2+u^2\) and \(2(t^2+u^2)^2\), respectively.

\Needspace{6\baselineskip}
\subsubsection{Prescribing the first exact level}\label{prescribing-the-first-exact-level}

\begin{paperresult}{Corollary 1H}{res-1H}
Fix a reduced real plane cubic \(X\) with real points dense in each geometric component, and suppose \(X\) is smooth or all its singularities are split real ordinary nodes. Every positive integer occurs as the first exact PPT extension level for a positive definite contraction quadratic on this same \(X\), simultaneously for every \(d\ge3\). In the smooth case there is a path \(g_\epsilon=z^2+\epsilon h\) with \(h\) positive definite and a strictly decreasing sequence \(\epsilon_s>0\) tending to zero such that level \(s\) is exact precisely for \(\epsilon\ge\epsilon_s\), with \(\epsilon_s\sim L^6/(4096b\,s^6)\), in the notation of \hyperref[s-3-3]{Section~3.3}. In the singular case there is a family \(g_c\), \(0\le c<1\), with first exact level \[
s_\star(g_c)=\left\lceil\frac{\pi}{3\arccos c}\right\rceil.
\] The matrices of this family are polynomial in \(c\) in suitable fixed coordinates, and rational parameters suffice to attain every level.

\end{paperresult}

\textbf{The three split-nodal cases.} For three nonconcurrent real lines, use their three fixed projective vertices as a basis. Give them unit length and mutual inner products \(c\in[0,1)\). The metric has eigenvalues \(1-c,1-c,1+2c\), and the angular cycle length in \hyperlink{res-1C}{Theorem~1C} is \(3\arccos c\). Together with the constructions above, this proves the formula in \hyperlink{res-1H}{Corollary~1H} on all singular cubics with only split real ordinary nodes and dense real points on each component. For \(k\ge2\), any \(c\) in the interval \[
\cos\big(\pi/(3(k-1))\big)<c\le\cos\big(\pi/(3k)\big)
\] has first exact level \(k\). The interval has nonempty interior and contains rational numbers. Level one is attained at \(c=0\). For example, \(c=3/4\), \(9/10\), and \(497/500\) give first exact levels \(2\), \(3\), and \(10\) on each type of cubic. The asymptotic \(\arccos c\sim\sqrt{2(1-c)}\) also gives the growth of this first level as \(g_c\) degenerates.

Theorems 1C--1G give the first exact level for every cubic covered by \hyperlink{res-1D}{Theorem~1D}, including infinity when the hierarchy is never exact. Three nonconcurrent real lines form a triangle on independent projective vertices and use \hyperlink{res-1C}{Theorem~1C}. The geometrically irreducible split-nodal and smooth cases use Theorems 1E and 1F, respectively, and the line and conic cases use \hyperlink{res-1G}{Theorem~1G}. In particular, this first level is independent of \(d\) throughout this class of cubics.

\section{Statements and Declarations}\label{statements-and-declarations}

\textbf{Data availability.} The ancillary files accompanying this preprint include the Python source and its JSON output. Run \texttt{python3 verify.py} from the ancillary directory. The formulas and tables provide the finite inputs.

\textbf{Use of AI tools.} ChatGPT was used to explore proof steps, draft and translate parts of the manuscript, and develop and check computational certificates. The mathematical arguments and the code are presented for verification in the article and its supplement.

\section{Appendix A. Explicit rational attainment at every finite level}\label{s-A}

\subsection{A.1 A signed evaluation construction}\label{s-A-1}

First take \(N=3\) and fix the real triangle \(X\) of coordinate lines in \(\mathbb P^2\), with cyclic vertices \(e_0,e_1,e_2\) and coordinate ring \(R=\mathbb R[x_0,x_1,x_2]/(x_0x_1x_2)\). On the edge from \(e_i\) to \(e_{i+1}\), indices taken modulo three, choose a real binary form \(p_i\) of degree \(d\) with \(d\) distinct real zeros away from the vertices and \[
p_i(e_i)=1,\qquad p_i(e_{i+1})=\varepsilon_i,\qquad
\varepsilon_i\in\{1,-1\},\quad \prod_i\varepsilon_i=-1 .
\] These forms can be chosen over \(\mathbb Q\). On an edge with coordinates \((t,u)\), take \(\prod_{j=1}^d(t-\lambda_j u)\) where the distinct nonzero rational roots have product of absolute value one, and choose its sign by negating a single root as needed. The squared restrictions \(p_i^2\) agree at the vertices and define a form \(f\in R_{2d}\). In fact for any positive degree, the restriction map from \(R\) identifies homogeneous forms with triples of binary forms agreeing at the vertices, by assigning their mixed monomials to the respective edges and their pure monomials to the vertices. In particular \(\dim R_j=3j\) for \(j\ge1\). The form \(f\) is nonnegative on \(X(\mathbb R)\).

Choose a rational linear form \(\ell\) nonzero at the vertices whose zero on each edge avoids the roots of \(p_i\). In the affine coordinate \(v\) with representatives \(v e_i+e_{i+1}\) on edge \(i\), denote the root of \(\ell\) by \(b_i\). The \(3D\) linear conditions on \(R_D\) given by values at the roots of the \(p_i\) and by derivatives of orders \(0,\ldots,s-1\) at \(b_i\) are independent. In fact, any form in the common kernel restricts on edge \(i\) to \(c_i\ell^s p_i\). Agreement at the vertices requires \(c_{i+1}=\varepsilon_i c_i\). The product of the signs is negative, so all the coefficients vanish.

For a small positive rational \(t\), let \(Z_t\) consist of the roots of the \(p_i\) and the \(s\) nodes \(b_i+jt\), \(1\le j\le s\), on each edge, with the indicated affine representatives. Fix any norm on \(R_D\). The sum of the squares of its evaluations satisfies \[
\sum_{z\in Z_t} a(z)^2\ \ge\ c\,t^{2(s-1)}\|a\|^2 \qquad(a\in R_D)
\] for a constant \(c>0\) and all sufficiently small \(t\). Indeed, replace the cluster evaluations by forward differences divided by \(t^k\), for \(0\le k<s\). These converge to the independent derivative conditions just proved. The transformed evaluation map thus has a uniform lower bound, and the norm of the transformation is at most a constant times \(t^{-(s-1)}\).

Choose rational linear forms \(\ell_1,\ell_2\) completing \(\ell\) to a basis, set \(q_t=\ell^2+t^2(\ell_1^2+\ell_2^2)\), and choose a real rational point representative \(w\) on an edge with \(\ell(w)f(w)\ne0\). Define \[
M_t=\sum_{z\in Z_t}\operatorname{ev}_z-t^{2s-1}\operatorname{ev}_w
\qquad\text{on }R_{2D}.
\] The evaluation bound shows that the middle form of \(M_t\) is positive definite on \(R_D\) for sufficiently small \(t\). At the roots of \(p_i\) the value of \(q_t^s f\) is zero. At the clustered nodes it is \(O(t^{2s})\), while \(q_t(w)^s f(w)\) tends to a positive number. Therefore \(M_t(q_t^s f)<0\) for all sufficiently small \(t\). The construction uses rational coefficients throughout and can be verified by testing positive definiteness and one strict inequality.

The functional \(\widetilde L(u)=M_t(q_t^s u)\) is positive on nonzero squares from \(R_d\). Indeed the product expansion contains the square \((\ell^s v)^2\) for a square \(v^2\), and \(\ell\) is a nonzero divisor in \(R\). Thus the middle rank of \(\widetilde L\) on the ambient degree-\(d\) forms is \(\dim R_d=3d\). Any separable representation of \(\widetilde L\) would consist of real point evaluations on \(X\). To see this, apply it to \((x_0x_1x_2\,v)^2\) for every \(v\) of degree \(d-3\). The squares all vanish under \(\widetilde L\), and a positive evaluation away from \(X\) gives a positive value for some \(v\). A representation on \(X\) would give \(\widetilde L(f)\ge0\), whereas \(\widetilde L(f)<0\). Therefore \(\widetilde L\) is entangled and has actual local rank three, since binary CS states are separable. An invertible real linear change of coordinates sends the multiplier \(q_t\) to the usual sum-of-squares form \(g\). Apply it to both functionals and normalize by the common trace. The resulting CS state has rank \(3d\) and a CS extension by \(s\) parties.

For \(N>3\), embed this ternary example and its extension into the prescribed local space and add positive coherent projectors along \(N-3\) complementary coordinate axes. Add the corresponding extended projectors at level \(D\), with the same weights. Both local and operator ranks add, and projection onto the original three-dimensional local support recovers the entangled ternary state up to scale. This independently proves attainment of \(3d+N-3\) in \hyperlink{res-1A}{Theorem~1A}.

\textbf{Rational three-qutrit examples.} Take \(d=3\), \(p_i(v,1)=(v-\tfrac12)(v-1)(v-2)\) on all three cyclic edges, \(\ell=x_0+x_1+x_2\), \((\ell_1,\ell_2)=(x_0,x_1)\), and \(w=e_1\). Let \(V\) be the square evaluation matrix with rows indexed by \(Z_t\) and columns indexed by the monomials of degree \(D\) with at least one zero exponent. Let \(u\) be the evaluation column at \(w\) in the same basis. If \(b\) solves \(V^{\mathsf T}b=u\), the middle form is positive definite exactly when \(V\) is invertible and \(t^{2s-1}b^{\mathsf T}b<1\). This follows by writing its matrix as \(V^{\mathsf T}(1-t^{2s-1}bb^{\mathsf T})V\). The following bounds use rational arithmetic. All four \(V\) are invertible, and all four marginals have rank nine.

\begin{longtable}[]{@{}
  >{\raggedleft\arraybackslash}p{(\columnwidth - 6\tabcolsep) * \real{0.2500}}
  >{\raggedleft\arraybackslash}p{(\columnwidth - 6\tabcolsep) * \real{0.2500}}
  >{\raggedleft\arraybackslash}p{(\columnwidth - 6\tabcolsep) * \real{0.2500}}
  >{\raggedleft\arraybackslash}p{(\columnwidth - 6\tabcolsep) * \real{0.2500}}@{}}
\toprule\noalign{}
\begin{minipage}[b]{\linewidth}\raggedleft
Additional copies \(s\)
\end{minipage} & \begin{minipage}[b]{\linewidth}\raggedleft
\(t\)
\end{minipage} & \begin{minipage}[b]{\linewidth}\raggedleft
Upper bound on \(t^{2s-1}b^{\mathsf T}b\)
\end{minipage} & \begin{minipage}[b]{\linewidth}\raggedleft
Upper bound on \(M_t(q_t^s f)\)
\end{minipage} \\
\midrule\noalign{}
\endhead
\bottomrule\noalign{}
\endlastfoot
1 & \(2^{-10}\) & \(1/100\) & \(-4/10^{4}\) \\
2 & \(2^{-14}\) & \(1/100\) & \(-1/10^{13}\) \\
3 & \(2^{-19}\) & \(1/100\) & \(-1/10^{29}\) \\
4 & \(2^{-25}\) & \(1/100\) & \(-5/10^{54}\) \\
\end{longtable}

\subsection{A.2 Additional exact examples of graph arrangements}\label{s-A-2}

\textbf{Rational equiangular examples.} The three vectors \(v_i=(1,1,1)+a e_i\) for rational \(a>0\) have a common norm and normalized pairwise inner products \[
c(a)=\frac{3+2a}{3+2a+a^2}.
\] They are independent and \(c(a)\) ranges continuously and strictly decreasingly from one to zero. The interval for every \(s_0\ge2\) can thus be met by rational triangle coordinates. Choosing the \(p_{ij}\) in the rational basis \(v_i\) gives common absolute endpoint values after unit normalization, which can be scaled out of the sphere argument. The separating positive definite functional can also be chosen rational, by density in the open set defined by positive definiteness and strict negativity on \(g^s f\). Here are sample rational parameters. The exact checks use \(\cos(k\theta)=T_k(c(a))\) and the Chebyshev recurrence \(T_{k+1}(c)=2cT_k(c)-T_{k-1}(c)\).

\begin{longtable}[]{@{}rrr@{}}
\toprule\noalign{}
\(a\) & \(c(a)\) & First exact level \(s_0\) \\
\midrule\noalign{}
\endhead
\bottomrule\noalign{}
\endlastfoot
\(3\) & \(1/2\) & 1 \\
\(1\) & \(5/6\) & 2 \\
\(2/3\) & \(39/43\) & 3 \\
\(1/2\) & \(16/17\) & 4 \\
\(1/7\) & \(161/162\) & 10 \\
\(1/19\) & \(1121/1122\) & 25 \\
\(1/77\) & \(17941/17942\) & 100 \\
\end{longtable}

\textbf{A rational triangle with three different angles.} Take the unit vertices \(u_0=(1,0,0)\), \(u_1=(3/5,4/5,0)\) and \(u_2=(4/5,0,3/5)\). The cosines on the cyclic edges are \(3/5,12/25,4/5\), and the angular length is \(\pi/2+\arccos(12/25)\), strictly between \(\pi/2\) and \(\pi\). Thus its first exact level is two. For \(d=3\) use the same cubic \(p\) and the same evaluation points as above in the new vertex basis. Give the nine root evaluations weight \(32\), the three differences weight \(1/81\), and the three vertices weight \(-4/5\). The degree-four middle matrix is positive definite by its leading principal minors, with determinant \(129370186317824/729\). For the witness \(f\) given by \(p^2\) on every edge, the evaluations give \(M(gf)=-4/25\) and \(M(g^4)=9180742429499/28125000\). This gives a rational rank-nine state separating levels one and two. The level-two witness certificate uses the same \(\Phi_i\) construction with the cosine \(c_{ij}\) of the particular edge. The matrix \(K\) with edge entries \(1-2c_{ij}^2\) and diagonal one is positive definite, with determinant \(36556/78125\). Since the vertices are unit vectors, the identity on the curve is \(g^2 f=\Phi^{\mathsf T}K\Phi\).

\textbf{A rational four-cycle.} In \(\mathbb R^4\) take vertices proportional to \((1,1,1,1)+\tfrac12 e_i\) and include just the four cyclic edges. The adjacent cosine is \(20/21\), with projective angle \(\theta\). We have \(0<2\theta<3\theta<\pi/2\), \(\cos(2\theta)=359/441>1/\sqrt2\), and \(\cos(3\theta)=5540/9261<1/\sqrt2\). These bounds can be checked by squaring. Thus \(8\theta<\pi<12\theta\), giving first exact level three by \hyperlink{res-1C}{Theorem~1C}. The degree-six Hankel cone on this four-cycle has entangled level-two marginals of operator rank twelve.

\textbf{A completion obstruction on \(K_4\).} Take the six vertex inner products in lexicographic edge order to be \(2/5,9/20,1/2,1/2,9/20,2/5\). With unit diagonal the vertex Gram matrix is positive definite, so this is a realizable geometry on independent vertices. Every edge angle is at least \(\pi/3\), and every cycle passes the first-level angular bound. At level one the all-negative signing forces a matrix whose quadratic form at the vector of ones is \(4-2\sum_{ij}\langle u_i,u_j\rangle=-7/5\). At level two all intervals contain zero since \(2\langle u_i,u_j\rangle^2-1<0\). The identity is feasible for every signing. The first exact level is therefore two.

\subsection{A.3 An explicit ambient witness for the triangle example}\label{s-A-3}

Use the vertex coordinates of \hyperref[s-1-1]{Section~1.1}, indices modulo three, and put \(p_i=p(y_i,y_{i+1})\). Choose the homogeneous representatives \[
f=\sum_i(p_i^2-y_i^6),\qquad
\Phi_i=p_i(y_i^2+\tfrac65 y_i y_{i+1})
       -p_{i-1}(y_i^2+\tfrac65 y_i y_{i-1})-y_i^5 .
\] These have the required restrictions on each edge. Write \(q=y_0y_1y_2\) and use the cubic and the three quadratics \[
\ell=\tfrac{49}{2}\sum_i y_i^3-\tfrac{35}{2}\sum_{i\ne j}y_i^2y_j,\qquad
r_i=-\tfrac{661}{120}y_i^2+\tfrac{191}{40}(y_{i+1}^2+y_{i-1}^2).
\] Set \(F=f+q\ell+600q^2\). Direct coefficient comparison, with \(\mathbf r=(r_i)_i\) and the matrix \(K\) of \hyperref[s-1-1]{Section~1.1}, gives \[
g^2F
=36(\Phi+q\mathbf r)^{\mathsf T}K(\Phi+q\mathbf r)+q^2(R+600g^2).
\] Here \(R\) is the symmetric quartic whose coefficient is constant on each of the following monomial types.

\begin{longtable}[]{@{}lr@{}}
\toprule\noalign{}
Type & Coefficient of each monomial in \(R\) \\
\midrule\noalign{}
\endhead
\bottomrule\noalign{}
\endlastfoot
\(y_i^4\) & \(60277/24\) \\
\(y_i^3y_j\) & \(673/6\) \\
\(y_i^2y_j^2\) & \(-63119/9\) \\
\(y_i^2y_jy_k\), with \(i,j,k\) distinct & \(-66878/3\) \\
\end{longtable}

Let \(a,b,c,d_0\) be the corresponding coefficients of \(R+600g^2\). In the basis \((y_i^2)_i,(y_{i+1}y_{i+2})_i\), take the following block Gram matrix. Each \(3\times3\) block has constant diagonal and constant off-diagonal entries.

\begin{longtable}[]{@{}lrr@{}}
\toprule\noalign{}
Block & Diagonal & Off-diagonal \\
\midrule\noalign{}
\endhead
\bottomrule\noalign{}
\endlastfoot
Upper-left & \(a\) & \(12100\) \\
Lower-right & \(c-24200\) & \(60500\) \\
Upper-right and lower-left & \(d_0/2-60500\) & \(b/2\) \\
\end{longtable}

The diagonal and off-diagonal coefficients yield the four monomial types above. On the span of the all-ones vector and on its orthogonal complement, the block matrix reduces to two \(2\times2\) matrices with positive first diagonal entries. Their determinants are \(3851659087/216\) and \(443816951/432\), respectively. Hence \(R+600g^2\) is a sum of squares and the displayed identity certifies ambient level-two separation. Because all evaluations defining \(\rho\) lie on \(q=0\), one has \(L_\rho(F)=-6/T\).

To compute the isotropic noise value, write \(x=Vy\) in physical coordinates, where the columns of \(V\) are the three \(v_i\) of \hyperref[s-1-1]{Section~1.1}. Then \(V^{\mathsf T}V=I+5J\) with \(J\) the all-ones matrix. If \(Z\) is a standard Gaussian vector in physical coordinates, \(Y=V^{-1}Z\) has covariance \(I-5J/16\). Radial integration gives \[
L_{\sigma_3}(F)=\frac{\mathbb E F(Y)}{\mathbb E\|Z\|^6}
=\frac{2113141}{286720}.
\] For the coefficient calculation one may pair the six factors in each monomial and multiply covariances within each pair, which follows by differentiating the Gaussian moment generating function. The radial denominator is \(3\cdot5\cdot7=105\). Substituting this value into the mixture witness gives the interval in \hyperref[s-2-3]{Section~2.3}.

\section{Appendix B. Reproducibility}\label{s-B}

Run the accompanying standard-library verifier with Python 3.10 or later using \texttt{python3\ verify.py}. It checks the three-qutrit extension and marginal ranks, cubic kernels, separation of consecutive levels, and the first-level witness on the same triangle, and the interpolation and the strict level-two Gram certificate for the smooth rank-nine example. It also verifies the signed-evaluation examples of \hyperref[s-A]{Appendix~A}, the endpoint identities, the graph examples, and the smooth and singular cubic certificates, including the prescribed first exact levels and the level-two split nodal specialization used in the signature argument. The ambient-witness check verifies the polynomial identity, the quartic Gram matrix, and the noise threshold. The rank and exactness arguments in Sections 2 and 3 are analytic proofs.

\end{document}